\documentclass[twocolumn,showpacs,pra,superscriptaddress,longbibliography]{revtex4-1}
\usepackage{times}
\usepackage{amsmath}
\usepackage{amssymb}
\usepackage{graphicx}
\usepackage[unicode]{hyperref}
\hypersetup{
   unicode=true,          
   plainpages=false,
   colorlinks=true,       
   citecolor=blue,        
}
\usepackage[caption=false,position=top,singlelinecheck=off,justification=raggedright]{subfig}
\usepackage{color}
\usepackage{pst-node}

\def \b1{{\bf 1}}

\newcommand{\bea}{\begin{eqnarray}}

\newcommand{\eea}{\end{eqnarray}}

\newcommand{\beq}{\begin{equation}}

\newcommand{\eeq}{\end{equation}}

\begin{document}
\title{{Parametrically} driven-dissipative three-level Dicke model}

\author{Jim Skulte}
\affiliation{Zentrum f\"ur Optische Quantentechnologien and Institut f\"ur Laser-Physik, Universit\"at Hamburg, 22761 Hamburg, Germany}
\affiliation{The Hamburg Center for Ultrafast Imaging, Luruper Chaussee 149, 22761 Hamburg, Germany}

\author{Phatthamon Kongkhambut}
\affiliation{Zentrum f\"ur Optische Quantentechnologien and Institut f\"ur Laser-Physik, Universit\"at Hamburg, 22761 Hamburg, Germany}

\author{Hans Ke{\ss}ler}
\affiliation{Zentrum f\"ur Optische Quantentechnologien and Institut f\"ur Laser-Physik, Universit\"at Hamburg, 22761 Hamburg, Germany}

\author{Andreas Hemmerich}
\affiliation{Zentrum f\"ur Optische Quantentechnologien and Institut f\"ur Laser-Physik, Universit\"at Hamburg, 22761 Hamburg, Germany}
\affiliation{The Hamburg Center for Ultrafast Imaging, Luruper Chaussee 149, 22761 Hamburg, Germany}

\author{Ludwig Mathey}
\affiliation{Zentrum f\"ur Optische Quantentechnologien and Institut f\"ur Laser-Physik, Universit\"at Hamburg, 22761 Hamburg, Germany}
\affiliation{The Hamburg Center for Ultrafast Imaging, Luruper Chaussee 149, 22761 Hamburg, Germany}

\author{Jayson G. Cosme}
\affiliation{National Institute of Physics, University of the Philippines, Diliman, Quezon City 1101, Philippines}

\date{\today}
\begin{abstract}

We investigate the three-level Dicke model, which describes a fundamental class of light-matter systems. We determine the phase diagram in the presence of dissipation, which we assume to derive from photon loss. Utilizing both analytical and numerical methods we characterize {the incommensurate time crystalline, light-induced, and light-enhanced superradiant states in the phase diagram for the parametrically driven system}. As a primary application, we demonstrate that a shaken atom-cavity system is naturally approximated via a parametrically driven-dissipative three-level Dicke model. 

\end{abstract}
\maketitle

\section{Introduction}

The Dicke model is a paradigmatic model capturing the physics of a fundamental class of light-matter systems \cite{Dicke}. 
The standard two-level Dicke model describes the interaction between $N$ two-level systems and a quantised single-mode light field. The dissipative or open standard Dicke model was first realized using an atom-cavity set-up allowing for an approximate description, in which the intra-cavity light field is adiabatically eliminated \cite{Baumann2010}. Later, it was also implemented in the recoil-resolved regime, which requires independent dynamical descriptions of the cavity and the matter field \cite{Klinder2015}. Meanwhile, extensions of the two-level Dicke models \cite{Hayn2011,Bastidas2012,Chitra2015,Zhiqiang2017,Soriente2018,Chiacchio2019,Buca2019,Stitely2020} and variations of the transversely pumped atom-cavity systems \cite{Habibian2013,Kollath2016,Mivehvar2017,Vaidya2018,Landini2018,Dogra2019,Bentsen2019,Jager2020} have been studied.

An important class of quantum optical phenomena derive from three-level systems interacting with light.
These phenomena include electromagnetically induced transparency (EIT) \cite{PhysRevLettBoller,RevModPhysFleischhauer} and lasing without inversion (LWI) \cite{PhysRevLettScully,Mompart_2000}, as well as methods such as stimulated Raman adiabatic passage (StiRAP) \cite{Gaubatz,RevModPhysVitanov}. They  are based primarily on three-level systems in a $\lambda$ or a V configuration. These three-level system configurations occur naturally in numerous physical systems, which is the origin of the universality of the phenomena that derive from them. In the context of the Dicke model, its generalisation to three-level atoms interacting with a multimode photonic field has been proposed in Ref.~\cite{Sung1979}. A similar three-level model has been used to demonstrate subradiance \cite{Crubellier1985,Crubellier1986,Cola2009,Wolf2018}.

\begin{figure}[!htb]
\centering
\includegraphics[width=1\columnwidth]{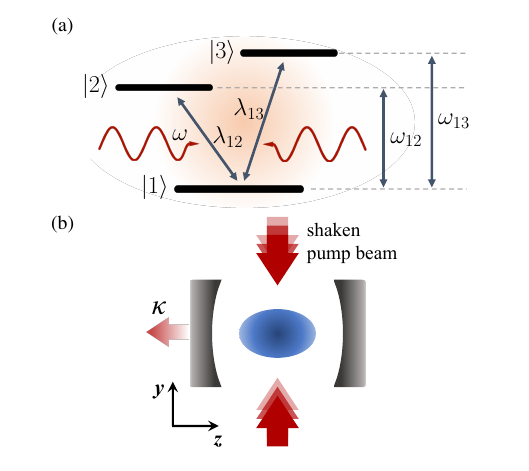}
\caption{(a) Three-level system coupled to a single light mode. (b) Schematic diagram of the shaken atom-cavity system. The cavity photon loss rate is $\kappa$. This atom-cavity configuration can emulate the driven-dissipative three-level Dicke model.}
\label{fig:schem} 
\end{figure}

In this work, we study a system of three-level atoms coupled to a photonic mode modelled by a three-level Dicke mode, in which the three-level system forms a V configuration as depicted in Fig.~\ref{fig:schem}(a). The three-level system can be described using pseudospin operators following the algebra of the SU(3) group. Our representation maps onto the standard SU(3) basis, the Gell-Mann matrices \cite{georgi2018lie}, spanning the Lie algebra in the defining representation of the SU(3) group. The Gell-Mann matrices are commonly used in particle physics to explain colour charges \cite{MARCIANO,Griffiths2008}. We obtain the equilibrium phase diagram of the three-level Dicke model in the presence of dissipation due to photon loss. Moreover, we show that periodic driving of the light-matter interaction strength may lead to the emergence of new nonequilibrium phases, such as an incommensurate time crystal (ITC), light-induced superradiance (LISR) and light-enhanced superradiance (LESR).

Here, we present a comprehensive discussion of {a parametrically} driven three-level Dicke model. We discuss its dynamical phase diagram including the incommensurate crystalline phase, predicted by us in Ref.~\cite{Cosme2019} and experimentally implemented in Ref.~\cite{joint}. We show that this phase is a characteristic signature of the driven three-level Dicke model. We give a detailed account on how this model can be approximately implemented by a light-driven atom-cavity system.

This work is organized as follows. In Sec.~\ref{sec:SU3}, we introduce the three-level Dicke model and discuss its phase diagram. We explore the dynamical phase diagram of the driven three-level Dicke model in Sect.~\ref{sec:DPD}. The mapping of a shaken atom cavity system onto the periodically driven three-level Dicke model is presented in Sec.~\ref{sec:PSU3}. In Sec.~\ref{sec:con}, we conclude this paper.

\section{Three-level Dicke Model}
\label{sec:SU3}
We are interested in the properties of the three-level Dicke model for a system of $N$ three-level atoms interacting with a quantised light mode, as schematically shown in Fig.~\ref{fig:schem}(a). Each atom has three energy states $|1\rangle$, $|2\rangle$, and $|3\rangle$. We define the three-level Dicke model by the Hamiltonian, 
\begin{align}\label{eq:ham}
H/ \hbar &= \omega \hat{a}^\dagger \hat{a} + \omega_{\mathrm{12}} \hat{J}^{12}_z   +\omega_{\mathrm{13}}\hat{J}^{\mathrm{13}}_z  \\ \nonumber
&+\frac{2}{\sqrt{N}}\left(\hat{a}^\dagger+\hat{a} \right) \left(\lambda_{12}\hat{J}^{\mathrm{12}}_x + \lambda_{13} \hat{J}^{\mathrm{13}}_x \right),
\end{align}
where $\omega$ is the photon frequency, $\omega_{nm}$ is the detuning between states $|m\rangle$ and $|n\rangle$, and $\lambda_{nm}$ is the light-matter interaction strength associated with the photon-mediated coupling between states $|n\rangle$ and $|m\rangle$. The bosonic operators $\hat{a}$ and $\hat{a}^\dagger$ annihilate and create a photon in the quantised light mode, respectively.  There are three classes of pseudospin operators $\hat{J}^{12}_\mu$, $\hat{J}^{13}_\mu$, and $\hat{J}^{23}_\nu$ with $\mu \in \{z,\pm \}$ and $\nu \in \{\pm \}$, corresponding to the transitions $|1\rangle \leftrightarrow |2\rangle$, $|1\rangle \leftrightarrow |3\rangle$, and $|2\rangle \leftrightarrow |3\rangle$, respectively. These operators obey the commutation relation of the SU(3) algebra (see Appendix \ref{ap:algebra}).
The $x$- and $y$-components of the pseudospins are defined as $\hat{J}^{\mathrm{\ell}}_x = (\hat{J}^{\mathrm{\ell}}_+ + \hat{J}^{\mathrm{\ell}}_- )/2$ and $\hat{J}^{\mathrm{\ell}}_y = (\hat{J}^{\mathrm{\ell}}_+ - \hat{J}^{\mathrm{\ell}}_- )/2i$, respectively with $\ell \in \{12,13,23 \}$.

Note that, in principle, there is a light-matter coupling term proportional to $\hat{J}^{\mathrm{23}}_x$ in Eq.~\eqref{eq:ham} \cite{Sung1979}. However, this term is neglected here since we are only interested in the case when $\omega_{12} \approx \omega_{13}$. This leads to a negligibly small $\lambda_{23}$  {since the light-matter coupling strength is proportional to the energy difference between the relevant states   \cite{sakurai_napolitano_2017,Baksic2013}}. Moreover, we could also use the Gell-Mann matrices as the representation of the SU(3) group in our system. 
To retain a form of the Hamiltonian reminiscent of the standard two-level Dicke model, which is often written using a representation of the SU(2) group, we instead use the pseudospin operators as described above. Nevertheless, the Gell-Mann matrices can be obtained from appropriate superpositions of the pseudospin operators (see Appendix \ref{ap:algebra}). 

The Hamiltonian in Eq.~\eqref{eq:ham} is, superficially, similar to the two-component Dicke model \cite{Landini2018,Chiacchio2019,Buca2019,Dogra2019} (see also Appendix \ref{ap:2comp} for a brief discussion). However, we emphasise that  unlike in the two-component Dicke model, which describes two types of two-level systems coupled through the light field, the pseudospin operators introduced in Eq.~\eqref{eq:ham} obey the SU(3) algebra resulting from the use of three-level systems. This fundamentally changes the dynamics of the {parametrically driven} system {out of equilibrium} since new terms corresponding to additional spin operators are now present in the equations of motion.

\begin{figure}[!htbp]
\centering
\includegraphics[width=1.0\columnwidth]{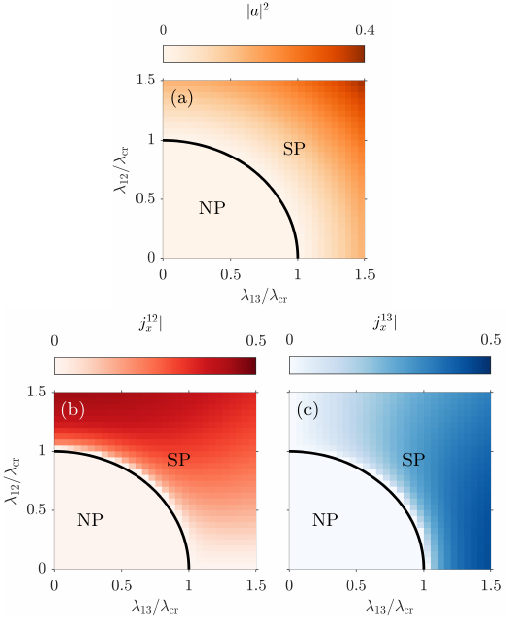}
\caption{Long-time average of the mean-field dynamics of the (a) cavity mode occupation $|a|^2$, (b) $|j^{12}_x|^2$, and (c) $|j^{13}_x|^2$ for $\omega = \omega_{12}=\omega_{13}=\kappa$. The black curve denotes the critical line separating the normal and superradiant phases in the thermodynamic limit.}
\label{fig:su3stat} 
\end{figure} 

\subsection{Holstein-Primakoff transformation}
\label{sec:HP}
{To obtain analytical predictions of the phase boundaries,} we employ a Holstein-Primakoff (HP) approximation in the thermodynamic limit, i.e. $N \rightarrow \infty$.
This leads to the following Hamiltonian
\begin{align}
\label{eq:HamHPSU3}
H/ \hbar &= \omega \hat{a}^\dagger \hat{a}+\omega_{12} \hat{a}^\dagger_{12} \hat{a}_{12}+\omega_{13} \hat{a}^\dagger_{13} \hat{a}_{13} \\ \nonumber &+\left(\hat{a}^\dagger+ \hat{a} \right) \left[ \lambda_{12} (\hat{a}^\dagger_{12}+\hat{a}_{12})+\lambda_{13} (\hat{a}^\dagger_{13}+\hat{a}_{13}) \right]~.
\end{align}
We obtain an elliptic equation for the critical light-matter coupling from the stability matrix (see Appendix \ref{ap:critlam} for details)
\begin{equation}\label{eq:crit}
\frac{\left( \kappa^2+\omega^2 \right)}{4 \omega}= \left( \frac{\lambda^2_{12}}{\omega_{12}}+\frac{\lambda^2_{13}}{\omega_{13}} \right).
\end{equation}
In the standard open Dicke model, $\lambda_{13}=0$, the critical light-matter coupling, $\lambda_\mathrm{cr}= \sqrt{(\kappa^2+\omega^2)\left(\omega_{12}/\omega\right)}/2$, is recovered \cite{Dimer2007}. To illustrate the resulting phases, we consider the case $\omega = \omega_{12} = \omega_{13}$. Then, the critical line in Eq.~\eqref{eq:crit} defines a circle in the parameter space spanned by $\lambda_{12}$ and $\lambda_{13}$, as seen in Fig.~\ref{fig:su3stat}. For combinations of light-matter coupling strengths $\{\lambda_{12},\lambda_{13}\}$ within the area enclosed by Eq.~\eqref{eq:crit}, the stable phase corresponds to a normal phase (NP), while those outside the area will lead to an instability towards the formation of a superradiant phase (SP).

\subsection{Phase diagram}\label{sec:eomstat}
{Next,} we employ a mean-field approximation $\langle \hat{a}  \hat{J}^\ell_\mu \rangle \approx\langle \hat{a} \rangle \langle \hat{J}^\ell_\mu \rangle $ {starting from Eq.~\eqref{eq:ham}} to obtain the dynamics of the system in a semiclassical approximation (see Appendix \ref{ap:eom} for details). This approximation becomes exact in the thermodynamic limit $N \rightarrow \infty$ in or near the steady state. Furthermore, we introduce the rescaled $c$-numbers $a \equiv \langle \hat{a} \rangle / \sqrt{N}$ and $j^\ell_\mu \equiv \langle \hat{J}^\ell_\mu \rangle /N$. The resulting {mean-field} equations of motion that we simulate are shown in Appendix~\ref{ap:mf_eom}.
 We further note that the SU(3) group inherits two Casimir charges, a quadratic $C_1$ and a cubic $C_2$. In contrast to this, the group SU(2) has only one quadratic Casimir charge, namely the total spin $J^2 =  (J_{x})^2 + (J_{y})^2 + (J_{z})^2$. The expressions for the charges are shown in Appendix~\ref{ap:algebra}. We track these quantities when solving the equations of motion to ensure convergence of our numerical results. In our simulations, we initialise in the normal phase $j^\ell_\mu =0$, except for $j^{12}_z=j^{13}_z=-1/2$.  This amounts to all the atoms initially occupying the lowest energy state $|1\rangle$. We initialise the cavity field as $a=10^{-2}$. 

An observable of interest is the occupation of the photonic mode $|a|^2$ as this differentiates the normal $(|a|^2 \rightarrow 0$ for $N \rightarrow \infty$
) and superradiant ($|a|^2>0$) phases.  Moreover, we are interested in the magnitude of the $x$-component of the collective spin operators corresponding to the transition $|1\rangle \leftrightarrow |2\rangle$ and $|1\rangle \leftrightarrow |3\rangle$, which are $| j^{12}_x |$ and $| j^{13}_x |$, respectively. 
In Fig.~\ref{fig:su3stat}, we present the long-time average of $|a|^2$, $| j^{12}_x |$ and $| j^{13}_x |$,  calculated from numerically solving the equations of motion.
Similar to the standard two-level Dicke model \cite{Kirton2019}, the photonic mode occupation or the $x$-component of the pseudospin operators can be regarded as order parameters, as they are zero in the NP and are nonzero in the SP. Furthermore, we demonstrate in Fig.~\ref{fig:su3stat} that the onset of superradiance according to our mean-field dynamics agrees with the analytical critical line defined by Eq.~\eqref{eq:crit}.
In the superradiant phase, $| j^{12}_x |>| j^{13}_x |$ for $\lambda_{12}>\lambda_{13}$ and $| j^{12}_x |<| j^{13}_x |$ for $\lambda_{12}<\lambda_{13}$, as inferred from Figs.~\ref{fig:su3stat}(b) and \ref{fig:su3stat}(c).

\begin{figure*}[!htpb]
\centering
\includegraphics[width=2.0\columnwidth]{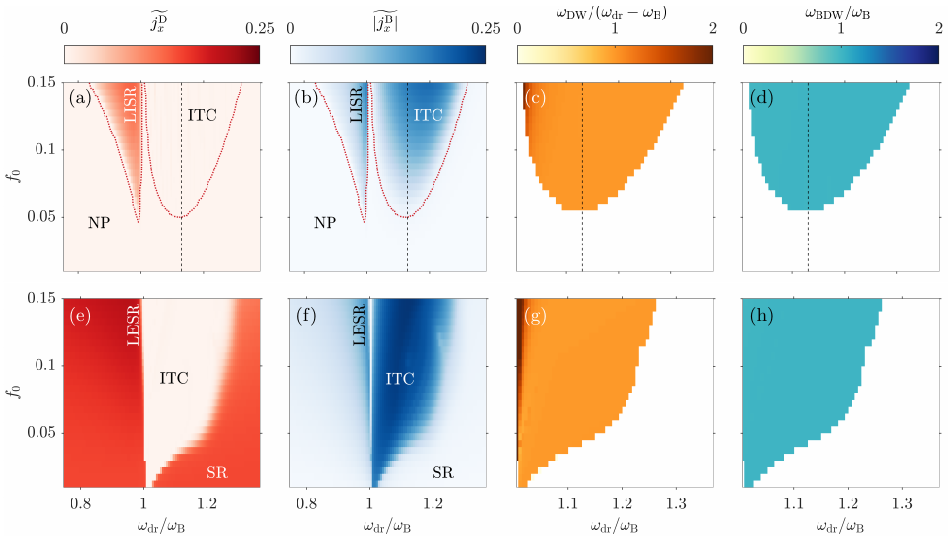}
\caption{Dynamical phase diagram for (a)-(d) $\lambda=0.98 \lambda_\mathrm{cr}$ and (e)-(f) $\lambda=1.02 \lambda_\mathrm{cr}$. Time-averaged (a), (e) $\widetilde{j^{\mathrm{D}}_x}$ and (b), (f) $\widetilde{|j^{\mathrm{B}}_x|}$ taken over 100 driving cycles, $\tau = 100T$, for varying modulation parameters. The dominant or peak frequency in the power spectrum of (c), (g) $j^{\mathrm{D}}_x$ and (d),(h) $j^{\mathrm{B}}_x$ for $\omega_{\mathrm{dr}}>\omega_{\mathrm{B}}$. The dotted lines in (a) and (b) denote the instability boundary according to the oscillator model. 
In (c), (d), (g), and (h), we are only showing the response frequencies $\omega_\mathrm{DW}$ and $\omega_\mathrm{BDW}$ for parameter sets, which yield $\widetilde{|j^{\mathrm{B}}_x|}>0.01$ and $\widetilde{j^{\mathrm{D}}_x}<0$.
Note that we are rescaling the response frequencies  in (c) and (g) to $\omega_\mathrm{DW}/(\omega_\mathrm{dr}-\omega_\mathrm{B})$ and it is rescaled in (d) and (h) to $\omega_\mathrm{BDW}/\omega_\mathrm{B}$. 
The vertical dashed line in (a)-(d) corresponds to the sum frequency $\omega_{\mathrm{sum}} = \omega_{\mathrm{LP}} + \omega_{\mathrm{B}} $.}
\label{fig:norm} 
\end{figure*}

\section{Parametrically Driven Open three-level Dicke model}
We now explore the parametrically driven three-level Dicke model by the Hamiltonian
\begin{align}\label{eq:ham2}
H/ \hbar &= \omega \hat{a}^\dagger \hat{a} +\omega_{\mathrm{D}}\hat{J}^{\mathrm{D}}_z   + \omega_{\mathrm{B}}\hat{J}^{\mathrm{B}}_z  
 +2 \phi(t) (\omega_{\mathrm{B}}-\omega_{\mathrm{D}})  \hat{J}^{\mathrm{DB}}_x  \\ \nonumber
&+\frac{2\lambda}{\sqrt{N}}\left(\hat{a}^\dagger+\hat{a} \right) \left(\hat{J}^{\mathrm{D}}_x -\phi(t) \hat{J}^{\mathrm{B}}_x \right).
\end{align}
{This particular choice of the Hamiltonian is motivated by its connection to the shaken atom-cavity system, which we will demonstrate and explore in more detail later.}
Comparing with the undriven case in Eq.~\eqref{eq:ham}, it can be seen that $\omega_{12} = \omega_\mathrm{D}$, $\omega_{13} = \omega_\mathrm{B}$, $\hat{J}^{12}_\mu= \hat{J}^\mathrm{D}_\mu$, $\hat{J}^{13}_\mu= \hat{J}^\mathrm{B}_\mu$, $\lambda_{12}=\lambda$. {We define $\phi(t)=f_0\sin(\omega_\mathrm{dr} t)$}, which then means that $\lambda_{13}=-f_0 \sin(\omega_\mathrm{dr} t)\lambda$. 
This labelling is motivated by the association of the pseudospins with the density wave states in the atom-cavity setup discussed later in Sec.~\ref{sec:PSU3}. 
For now, we simply note that the photonic mode corresponds to a single cavity mode while the operators $\hat{J}^\mathrm{D}_\mu$ and $\hat{J}^\mathrm{B}_\mu$ are associated with the density wave (DW) and bond-density wave (BDW) states in the shaken atom-cavity system, respectively \cite{Cosme2019}. A small term proportional to $\hat{J}^{\mathrm{23}}_x \equiv \hat{J}^{\mathrm{DB}}_x$ is included in Eq.~\eqref{eq:ham2} { since this will appear later when we show how the atom-cavity system can be mapped onto the specific form of the parametrically driven three-level Dicke model Eq.~\eqref{eq:ham2}}.

\label{sec:DPD}
\subsection{Holstein-Primakoff transformation}

In Sect.~\ref{sec:HP}, we have applied the HP transformation to the undriven system described by Eq.~\eqref{eq:HamHPSU3}. We now extend this analysis to include the driving term. Applying the transformation and identifying $\hat{d} \equiv \hat{a}_\mathrm{12}$ and $ \hat{b}\equiv \hat{a}_\mathrm{13}$, we obtain a HP Hamiltonian shown in Eq.~\eqref{eq:HamHPdriven} of Appendix~\ref{ap:hp}. {In particular, we are interested in $d \equiv \langle \hat{d} \rangle$ and $b \equiv \langle \hat{b} \rangle$.}

We recall that for a quantum harmonic oscillator $\hat{f}^\dagger=\sqrt{\omega_F/ \hbar}(x_F-(i/\omega_F)p_F)$ and $\hat{f}=\sqrt{\omega_F/ \hbar}(x_F+(i/\omega_F)p_F)$. Then, we can express the corresponding HP Hamiltonian in momentum-position representation as
\begin{align}\label{eq:hamhpxp}
H &= \frac{\omega^2}{2}\hat{x}^2+\frac{\hat{p}^2}{2}+\frac{\omega_\mathrm{D}^2}{2}\hat{x}^2_\mathrm{D}+\frac{\hat{p}_\mathrm{D}^2}{2}+\frac{\omega_\mathrm{B}^2}{2}\hat{x}^2_\mathrm{B}+\frac{\hat{p}_\mathrm{B}^2}{2} \\ \nonumber &+ 2\lambda \sqrt{\omega \omega_\mathrm{D}}\hat{x} \hat{x}_\mathrm{D}-2 \phi(t)\lambda \sqrt{\omega \omega_\mathrm{B}}\hat{x} \hat{x}_\mathrm{B}\\ \nonumber &+\phi(t)(\omega_{\mathrm{B}}-\omega_{\mathrm{D}})\sqrt{\omega_\mathrm{D} \omega_\mathrm{B}}  \left( \hat{x}_\mathrm{D}\hat{x}_\mathrm{B}+\frac{\hat{p}_\mathrm{D}\hat{p}_\mathrm{B}}{\omega_\mathrm{D} \omega_\mathrm{B}} \right).
\end{align}
This has the form a Hamiltonian for three coupled oscillators:  (i) \textit{cavity oscillator}, (ii) \textit{DW oscillator}, and (iii) \textit{BDW oscillator} with frequencies $\omega$,  $\omega_\mathrm{D}$,  and $\omega_\mathrm{B}$, respectively.
Here, the two coupling constants connecting the BDW oscillator to the cavity and DW oscillators are periodically switched on and off or parametrically driven. Interestingly, due to the shaking of the pump, the momenta of the DW and BDW oscillators are also periodically coupled as seen from the last term in Eq.~\eqref{eq:hamhpxp}. {However, we find that this does not alter the qualitative features of the dynamics, as shown in Fig.~\ref{fig:w_equal} in the Appendix \ref{ap:mf_eom}.}

{We initialise the system in the normal state corresponding to having $d=0$ and $b=0$, which amounts to the absence of bosons in the excited states $|2\rangle$ and $|3\rangle$, respectively. Note that a small non-zero occupation of the photonic mode $\langle \hat{a} \rangle \equiv a = 10^{-2}$ is necessary to push the system out of the normal phase when it becomes an unstable state \cite{Chiacchio2019}.
The dynamics is obtained according to Eq.~\eqref{eq:eomhpdriven} for varying driving strength $f_0$ and driving frequency $\omega_{\mathrm{dr}}$.}  A parametric resonance in a linear system corresponding to a bilinear Hamiltonian, such as the simplified toy model Eq.~\eqref{eq:HamHPdriven}, manifests itself as an oscillatory solution with exponentially diverging amplitude. The dotted curves in  Figs.~\ref{fig:norm}(a)-(d) denote the points in the $(\omega_{\mathrm{dr}},f_0)$-space, where $(b+b^*)$ exceeds unity within the first 100 driving cycles, signalling a diverging solution  (see also Fig.~\ref{fig:npsupcav}). They indicate the regions where the normal phase is unstable towards a different collective phase. 

We identify two resonances responsible for the driving-induced destabilisation of the normal phase: (i) resonance at the BDW oscillator frequency $\omega_\mathrm{B}$ and (ii) a sum resonance involving $\omega_\mathrm{B}$ and the lower polariton frequency $\omega_\mathrm{LP}$ of the {atomic modes dressed by the cavity mode forming the lower polariton state \cite{Mivehvar2021}}. Note that we derive the expression for $\omega_\mathrm{LP}$ within the HP approach and we describe our method for obtaining the lower polariton frequency by exploiting a parametric resonance in Appendix~\ref{ap:lp}. The resonance frequencies are identified as the driving frequencies with the lowest modulation strength needed to induce an exponential instability. For $\omega_{\mathrm{dr}} < \omega_{\mathrm{B}}$, the resonance frequency is close to $\omega_\mathrm{B}$.  For $\omega_{\mathrm{dr}} > \omega_{\mathrm{B}}$, the sum resonance at $\omega_\mathrm{sum} = \omega_\mathrm{B} + \omega_\mathrm{LP}$ is the main mechanism, as highlighted by the vertical dashed line in Figs.~\ref{fig:norm}(a)-(d) (see also Fig.~\ref{fig:npsupcav}).

\subsection{Dynamical phase diagrams}

To further understand the resonant collective phases, we obtain the dynamics of the system. Within the mean-field approximation, we simulate the semiclassical equations of motion shown in Appendix \ref{ap:mf_eom}.
{Similar to the HP theory in the previous subsection, we initialise the system in the normal phase with small non-zero} occupation of the photonic mode $a = 10^{-2}$, We further choose $j^\ell_\mu =0$, except for $j^\mathrm{D}_z=j^\mathrm{B}_z=-1/2$. 
In addition to the photonic mode occupation $|a|^2$, we are also interested in the $x$-component of the pseudospins $ j^\mathrm{D}_x$ and  $j^\mathrm{B}_x$. 
Time is in units of the modulation period $T=2\pi/\omega_\mathrm{dr}$. The parameters for the simulation are shown in Appendix \ref{ap:param}.

In Fig.~\ref{fig:npsrdyn}, we present exemplary dynamics for resonant modulation specifcially for $\omega_{\mathrm{dr}}=1.05\,\omega_{\mathrm{B}}$. We choose the light-matter coupling strengths close to the phase boundary between the normal and superradiant phases, specifically $\lambda = 0.98\,\lambda_{\mathrm{cr}}$ and $\lambda = 1.02\,\lambda_{\mathrm{cr}}$, respectively. In the absence of driving, $f_0=0$, we reproduce the prediction of a  normal phase NP and superradiant phase  SP from the standard two-level Dicke model.  Periodic driving closed to but blue-detuned from $\omega_\mathrm{B}$ leads to similar long-time behaviour for $\lambda < \lambda_\mathrm{cr}$ and $\lambda > \lambda_\mathrm{cr}$. That is, the spin components related to the order parameters in the atom-cavity system, $j^{\mathrm{D}}_x$ and $j^{\mathrm{B}}_x$, periodically changes their sign concomitant to pulses of light being emitted. The slow subharmonic oscillations in $j^{\mathrm{D}}_x$, as exemplified in Fig.~\ref{fig:npsrdyn}(b), reflects the temporal periodicity of the entire light-matter system. 
{Note that}, $j^{\mathrm{B}}_x$ rapidly switches sign, as shown in Figs.~\ref{fig:npsrdyn}(c) and \ref{fig:npsrdyn}(f). 
\begin{figure*}[!htpb]
\centering
\includegraphics[width=2.0\columnwidth]{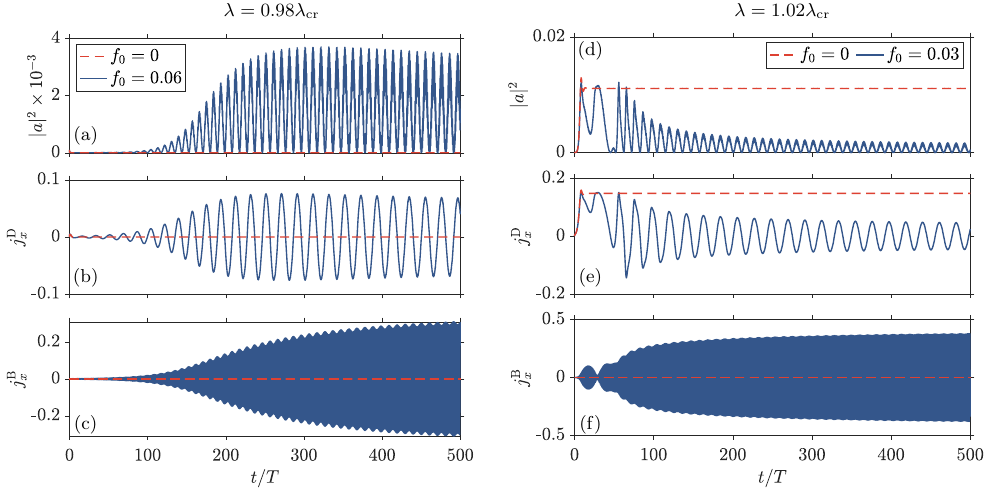}
\caption{Comparison between unmodulated and resonantly modulated dynamics for light-matter coupling strengths close to the NP-SR phase of the unmodulated system, (a)-(c) $\lambda = 0.98 \lambda_{\mathrm{cr}}$ and (d)-(f) $\lambda = 1.02 \lambda_{\mathrm{cr}}$. The relevant observables are the (a),(d) cavity mode occupation $|a|^2$, and the order parameters (b),(e) $j^{\mathrm{D}}_x$ and (c),(f) $j^{\mathrm{B}}_x$. The modulation frequency is fixed at $\omega_{\mathrm{dr}}=1.05\omega_{\mathrm{B}}$.}
\label{fig:npsrdyn} 
\end{figure*} 
We quantify the dynamical regimes in the system using the response frequencies $\omega_{\mathrm{DW}}$ and $\omega_{\mathrm{BDW}}$, which we define as the frequency at which $j^{\mathrm{D}}_x$ and $j^{\mathrm{B}}_x$ has a maximum {in the power spectrum}. Considering blue-detuned driving with respect to the BDW oscillator frequency $\omega_\mathrm{dr} > \omega_\mathrm{B}$, we find that the DBDW phase is characterised by fast oscillations of $j^{\mathrm{B}}_x$ at $\omega_{\mathrm{BDW}}=\omega_{\mathrm{B}}$ and slow oscillations of $j^{\mathrm{D}}_x$ at $\omega_{\mathrm{DW}} = \omega_{\mathrm{dr}}-\omega_{\mathrm{B}}$.
These observations are valid for both $\lambda < \lambda_\mathrm{cr}$ and $\lambda > \lambda_\mathrm{cr}$, as demonstrated in Figs.~\ref{fig:norm}(c), \ref{fig:norm}(d), \ref{fig:norm}(g), and \ref{fig:norm}(h), where the relations $\omega_{\mathrm{DW}}/(\omega_{\mathrm{dr}}-\omega_{\mathrm{B}})=1$ and $\omega_{\mathrm{BDW}}/\omega_{\mathrm{B}}=1$ are satisfied over a wide range modulation parameters. In general, the system's response frequency $\omega_{\mathrm{DW}}$ is subharmonic and incommensurate with respect to the driving frequency $\omega_\mathrm{dr}$, underpinning the classification of the DBDW phase as an ITC.
Thus, we show that the emergence of the ITC phase is one of the signatures of the parametrically driven three-level Dicke model. In contrast, the system has a harmonic response, meaning that $|a|^2$ and $j^{\mathrm{D}}_x$ have the same response frequency $\omega_{\mathrm{DW}} = 2\omega_{\mathrm{dr}}$ \cite{Cosme2019}, for combinations of driving parameters outside the dark areas in Figs.~\ref{fig:norm}(c), \ref{fig:norm}(d), \ref{fig:norm}(g), and \ref{fig:norm}(h), including red-detuned driving $\omega_\mathrm{dr} < \omega_\mathrm{B}$. 

In the ITC phase for $\omega_\mathrm{dr}>\omega_\mathrm{B}$, the oscillations of $j^{\mathrm{D}}_x$ and $j^{\mathrm{B}}_x$ around zero translate to vanishing time-averaged values,
\begin{equation}
\widetilde{j^\mathrm{\ell}_x} = \frac{1}{\tau}\int_0^\tau j^\mathrm{\ell}_x\, dt.
\end{equation}
This property is visible in the light area in Fig.~\ref{fig:norm}(e). Note, however, that even though $\widetilde{j^{\mathrm{D}}_x} = 0$, the time-averaged cavity mode occupation $\widetilde{|a|^2}$ does not necessarily vanish, especially when ${j^{\mathrm{D}}_x} $ has nonzero oscillation amplitude, as shown in Figs.~\ref{fig:norm}(a) and \ref{fig:npsupcav}(a).
The normal phase has ${j^{\mathrm{D}}_x} = 0$ for all times and as such, $\widetilde{|j^{\mathrm{D}}_x|} $ also vanishes, albeit trivially, similar to the ITC phase. Therefore, to distinguish between the normal phase and the ITC phase, we calculate  $\widetilde{|j^{\mathrm{B}}_x|} $, a quantity that vanishes for the normal phase and is nonzero for the ITC phase. 
In Figs.~\ref{fig:norm}(b) and \ref{fig:norm}(f), it can be seen that the BDW states are resonantly excited not only for the ITC phase in $\omega_\mathrm{dr} > \omega_\mathrm{B}$ but also for red-detuned driving $\omega_\mathrm{dr} < \omega_\mathrm{B}$. We emphasise that the dynamical response for $\omega_\mathrm{dr} < \omega_\mathrm{B}$ remains harmonic, making this phase distinct from the ITC, normal, and superradiant phases.

We now focus on red-detuned driving $\omega_\mathrm{dr} < \omega_\mathrm{B}$ to illustrate the effects of resonantly exciting the BDW states in this case. For $\lambda<\lambda_\mathrm{cr}$, the normal phase, expected to be dominant in the absence of driving, is suppressed, which then gives rise to a superradiant phase enabled by the excitation of the BDW states. We call this resonant phase for $\lambda<\lambda_\mathrm{cr}$ and $\omega_\mathrm{dr} < \omega_\mathrm{B}$ the \textit{light-induced superradiant} (LISR) phase. In this phase, the long-time average of the cavity mode occupation $|a|^2$ and $j^{\mathrm{D}}_x$ are both nonzero, similar to the superradiant phase, as seen from the resonance lobe in Figs.~\ref{fig:norm}(a) and \ref{fig:npsupcav}(a) for $\omega_\mathrm{dr}<\omega_\mathrm{B}$. However, the occupation of BDW states, demonstrated in Fig.~\ref{fig:norm}(b), distinguishes the LISR phase from the usual SR phase in the undriven case.
An analogous effect for $\lambda>\lambda_\mathrm{cr}$ is the enhancement of the superradiant phase, the stationary phase in the absence of driving. This \textit{light-enhanced superradiant} (LESR) phase is identified by an increase in $|a|^2$ and $j^{\mathrm{D}}_x$, accompanied by large amplitude oscillations of $j^{\mathrm{B}}_x$, as shown in Figs.~\ref{fig:npsupcav}(b), \ref{fig:norm}(e), and  \ref{fig:norm}(f). In addition to the ITC phase, the presence of LISR and LESR phases, depending on $\lambda$, is another signature of the driven-dissipative three-level Dicke model.

\section{Emulation using a Shaken Atom-Cavity system}
\label{sec:PSU3}
We now show that the parametrically driven open three-level Dicke model can be emulated by a shaken atom-cavity system. To this end, we first describe the many-body Hamiltonian of the shaken atom-cavity. Then, we present the approximation needed to obtain Eq.~\eqref{eq:ham2} from the atom-cavity Hamitlonian.

\subsection{Shaken atom-cavity Hamiltonian}

We consider a minimal model for describing the dynamics along the pump and cavity directions of an atom-cavity system schematically depicted in Fig.~\ref{fig:schem}(b).
The corresponding many-body Hamiltonian is given by \cite{Cosme2019}
\begin{align}\label{eq:hamac}
\hat{H}&/\hbar = - \delta_{\mathrm{C}} \hat{a}^\dagger \hat{a} + \int dy dz \hat{\Psi}^\dagger (y,z) \biggl[ -\frac{\hbar}{2m} \nabla^2 \\ \nonumber
&-\omega_\mathrm{rec}\epsilon_\mathrm{p} \cos^2(ky+\phi(t)) + U_0  \hat{a}^\dagger \hat{a} \cos^2(kz)  \\ \nonumber
&-\sqrt{\omega_\mathrm{rec}|U_0|\epsilon_\mathrm{p}}\cos(ky+\phi(t))\cos(kz)(a^\dagger+a)  \biggr] \hat{\Psi}(y,z),
\end{align}
where $\hat{a}$ ($\hat{a}^{\dagger}$) annihilates (creates) a photon in the single-mode cavity and $\hat{\Psi}(y,z)$ is the bosonic field operator for the atoms with mass $m$.
The pump-cavity detuning is  $\delta_{\mathrm{C}}$. The frequency shift per atom is taken to be red-shifted $U_0<0$. The pump intensity $\epsilon_{\mathrm{p}}$ is measured in units of the recoil energy $E_{\mathrm{rec}}=\hbar^2 k^2/2m$, where the wavevector is $k=2\pi/\lambda_\mathrm{p}$. Note that in Eq.~\eqref{eq:hamac}, we neglect the effects of short-range collisional interaction. The pump lattice is periodically shaken by introducing a time-dependent phase in the pump mode 
\begin{equation}
\phi(t) = f_0 \sin(\omega_{\mathrm{dr}} t),
\end{equation}
where $f_0$ is the unitless modulation strength and $\omega_{\mathrm{dr}}$ is the modulation frequency. The characteristic timescale is thus set by the driving period $T=2\pi/\omega_{\mathrm{dr}}$.

The dynamics of the atom-cavity system follows from the Heisenberg-Langevin equations \cite{Ritsch2013,Mivehvar2021},
\begin{align}
\frac{\partial}{\partial t} \hat{\Psi} &= \frac{i}{\hbar}[\hat{H}, \hat{\Psi}] \\
\frac{\partial}{\partial t} \hat{a} &= \frac{i}{\hbar}[\hat{H}, \hat{a}] - \kappa \hat{a} + \xi
\end{align}
where $\kappa$ is the cavity dissipation rate and the associated fluctuations are captured by the noise term $\xi$ satisfying $\langle \xi^*(t)\xi(t') \rangle = \kappa \delta(t-t')$. In the mean-field limit of large particle number $N$, quantum fluctuations are neglected and the bosonic operators can be approximated as $c$-numbers. The dynamics can then be obtained by numerically solving the resulting coupled differential equations corresponding to the equations of motion of the system. This approach and its extension beyond a mean-field approximation have been successfully used to predict and observe various dynamical phases in the transversely pumped atom-cavity system from a driving-induced renormalisation of the phase boundary to time crystals \cite{Cosme2018,Georges2018,Cosme2019,Kessler2019,Kessler2020,Kessler2021,Georges2021}.

\subsection{Low-momenta approximation}

The atom-cavity system can be mapped onto the Dicke model using a low-momenta approximation. To this end, we assume that the majority of the atoms only occupy the five-lowest momentum modes, namely the zero-momentum mode, $|k_y,k_z\rangle =  |0,0\rangle$, and the states associated with the self-organised chequerboard phase, $|\pm k, \pm k\rangle$. These momentum modes are coupled by the scattering of photons between the pump and cavity fields. This low-momenta approximation is valid close to the phase boundary between the homogeneous BEC phase and the self-organised DW phase. 

Resonant shaking has been shown to lead to the emergence of an incommensurate time crystal, where atoms localise at positions between the antinodes of the pump lattice \cite{Cosme2019,joint}. That is, in addition to the spatial mode $\cos(ky)\cos(kz)$ in the DW phase, the atoms are driven into additional states, namely the BDW states, as the atomic distribution acquires an overlap with the spatial mode $\sin(ky)\cos(kz)$. 
Note that this mode is made available by the periodic shaking of the pump lattice since it explicitly breaks the spatial symmetry along the pump axis. 
Owing to how the system periodically switches between superpositions of DW and BDW states, we call this dynamical phase as the dynamical BDW (DBDW) phase. Since the DBDW phase has been previously identified as an incommensurate time crystal (ITC), we will use the term DBDW and ITC phase interchangeably.

The atomic field operator is expanded to include the relevant spatial modes
\begin{equation}\label{eq:expa}
\hat{\Psi}(y,z) = \hat{c}_1 +2~\hat{c}_2  \cos(ky)\cos(kz) +2~\hat{c}_3 \sin(ky)\cos(kz),
\end{equation}
where the $c_i$ are bosonic annihilation and creation operator.
We use this expansion in the  many-body Hamiltonian Eq.~\eqref{eq:hamac}. Evaluating the integrals within one unit cell and for weak driving $f_0 \ll 1$, we obtain a Hamiltonian in a reduced subspace 
\begin{align}\label{eq:lowm}
&H/\hbar =-\delta_\mathrm{C} \hat{a}^\dagger \hat{a}+ 2 \omega_\mathrm{rec}(\hat{c}^\dagger_2 \hat{c}_2+\hat{c}^\dagger_3\hat{c}_3) + \frac{U_0}{2}\hat{a}^\dagger \hat{a}\biggl[ \hat{c}^\dagger_1 \hat{c}_1  \\ \nonumber
&+ \frac{3}{2}( \hat{c}^\dagger_2 \hat{c}_2+ \hat{c}^\dagger_3 \hat{c}_3) \biggr] -\frac{\omega_\mathrm{rec}\epsilon_\mathrm{p}}{4}\biggl[2(\hat{c}^\dagger_1\hat{c}_1+\hat{c}^\dagger_2\hat{c}_2+\hat{c}^\dagger_3\hat{c}_3) \\ \nonumber
&+ (\hat{c}^\dagger_2\hat{c}_2-\hat{c}^\dagger_3\hat{c}_3) - 2\phi(t)(\hat{c}^\dagger_2\hat{c}_3+\hat{c}^\dagger_3\hat{c}_2)  \biggr] -\frac{\sqrt{\omega_\mathrm{rec}|U_0|\epsilon_\mathrm{p}}}{2}  \\ \nonumber
&\qquad \times (\hat{a}^\dagger+\hat{a})  \biggl[ (\hat{c}^\dagger_1\hat{c}_2+\hat{c}^\dagger_2\hat{c}_1) -\phi(t)(\hat{c}^\dagger_1\hat{c}_3+\hat{c}^\dagger_3\hat{c}_1) \biggr].
\end{align}

\subsection{Schwinger boson representation}

We transform the bosonic operators in Eq.~\eqref{eq:lowm} into collective pseudospin operators through the Schwinger boson representation. The additional spatial mode $\sin(ky)\cos(kz)$ is described by the operator ${c}_3$, so the atomic motion is represented as a three-level system. We introduce the pseudospin operators obeying SU(3) algebra via
\begin{align}\label{eq:hols}
N&= \hat{c}^\dagger_1 \hat{c}_1+ \hat{c}^\dagger_2 \hat{c}_2+ \hat{c}^\dagger_3 \hat{c}_3 \\ \nonumber
\hat{J}^\mathrm{D}_+&= \hat{c}^\dagger_2 \hat{c}_1, \quad \hat{J}^\mathrm{D}_-= \hat{c}^\dagger_1 \hat{c}_2, \quad \hat{J}^\mathrm{D}_z = \frac{1}{2}\left(  \hat{c}^\dagger_2 \hat{c}_2-  \hat{c}^\dagger_3 \hat{c}_3-\hat{c}^\dagger_1 \hat{c}_1\right), \\ \nonumber
\hat{J}^\mathrm{B}_+ &= \hat{c}^\dagger_3 \hat{c}_1, \quad \hat{J}^\mathrm{B}_- = \hat{c}^\dagger_1 \hat{c}_3, \quad \hat{J}^\mathrm{B}_z = \frac{1}{2}\left(  \hat{c}^\dagger_3 \hat{c}_3-  \hat{c}^\dagger_2 \hat{c}_2-\hat{c}^\dagger_1 \hat{c}_1 \right),
 \\ \nonumber
\hat{J}^\mathrm{DB}_+&= \hat{c}^\dagger_2 \hat{c}_3, \quad \hat{J}^\mathrm{DB}_-= \hat{c}^\dagger_3 \hat{c}_2.
\end{align}
This representation suggests that the operators $\hat{J}^\mathrm{D}_\mu$ are associated with the DW state while $\hat{J}^\mathrm{B}_\mu$ are related to the BDW state.
Applying the commutation relations for the bosonic operators $[\hat{c}_m,\hat{c}^{\dagger}_n] = \delta_{mn}$,  we recover the same commutation relations for the pseudospin operators presented in Eq.~\eqref{eq:comm}. That is, we identify $\hat{J}^\mathrm{D}_\mu \equiv \hat{J}^\mathrm{12}_\mu$, $\hat{J}^\mathrm{B}_\mu \equiv \hat{J}^\mathrm{13}_\mu$, and $\hat{J}^\mathrm{DB}_\mu \equiv \hat{J}^\mathrm{23}_\mu$. 

Substituting the Schwinger boson representation in Eq.~\eqref{eq:hols} into Eq.~\eqref{eq:lowm} yields the driven-dissipative three-level Dicke model Eq.~\eqref{eq:ham2}.
Within the shaken-atom cavity platform, the effective cavity field frequency is $\omega = (3U_0 N)/4-\delta_\mathrm{C} = U_0 N/4 - \delta_\mathrm{eff}$, the effective pump-cavity detuning is $\delta_\mathrm{eff}$, and the light-matter coupling strength is $\lambda / \sqrt{N}=-\sqrt{\omega_\mathrm{rec}\epsilon_\mathrm{p} |U_0|}/{2}$. The pump intensity $\epsilon_\mathrm{p}$ shifts the frequencies of the pair of two-level transitions, $\omega_{\mathrm{D}} = 2 \omega_\mathrm{rec}(1-\epsilon_\mathrm{p}/8)$ and $\omega_{\mathrm{B}} = 2 \omega_\mathrm{rec}(1+\epsilon_\mathrm{p}/8)$. 
We can infer from Eq.~\eqref{eq:ham2} that weak periodic shaking effectively leads to a parametric driving of the light-matter coupling between the cavity and the spin associated to the BDW state. With these correspondences, we find that indeed the shaken atom-cavity system can be approximated by the driven three-level Dicke model presented in Eq.~\eqref{eq:ham2} and discussed in Sect.~\ref{sec:DPD}. Moreover, we can identify the order parameters of the self-organised density wave states, namely the DW order parameter $\Theta_\mathrm{DW} = \langle \cos(ky)\cos(kz) \rangle =  j^\mathrm{D}_x$ and the BDW order parameter $\Theta_\mathrm{BDW} =  \langle \sin(ky)\cos(kz) \rangle = j^\mathrm{B}_x$. 

\subsection{Comparison with the full atom-cavity model}\label{ap:comparison}

\begin{figure}[!htpb]
\centering
\includegraphics[width=1.0\columnwidth]{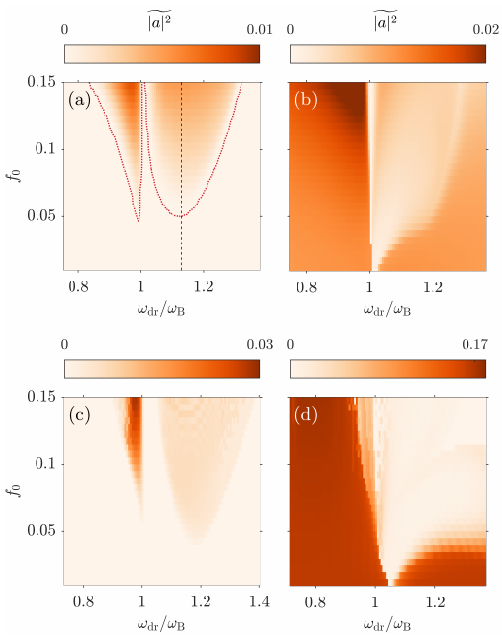}
\caption{Time-averaged cavity mode occupation $|a|^2$ taken over 100 driving cycles, $\tau = 100T$, according to (a),(b) the three-level Dicke model and (c),(d) the full atom-cavity model. For the three-level model, the light-matter coupling strengths are (a) $\lambda = 0.98 \lambda_{\mathrm{cr}}$ and (b) $\lambda =1.02 \lambda_{\mathrm{cr}}$. The broken lines denote the instability boundary from the oscillator model. The vertical dashed line in (a) corresponds to the sum frequency $\omega_{\mathrm{sum}} = \omega_{\mathrm{LP}} + \omega_{\mathrm{B}} $ involving the lower polariton frequency $\omega_{\mathrm{LP}}$. $\omega_{\mathrm{LP}}$ has the value $\omega_{\mathrm{LP}}/2\pi \approx 1.06~\mathrm{kHz}$ for this example. For the full atom-cavity model, the pump strengths are (c) $\epsilon_{\mathrm{p}}=0.96\epsilon_{\mathrm{cr}}$ and (d) $\epsilon_{\mathrm{p}} =1.04\epsilon_{\mathrm{cr}}$, which corresponds to the $\lambda =0.98\lambda_\mathrm{cr}$ and $\lambda =1.02 \lambda_\mathrm{cr}$, respectively.}
\label{fig:npsupcav} 
\end{figure}

We compare the dynamics of the cavity mode occupation and the DW order parameter for the full atom-cavity model Eq.~\eqref{eq:hamac} and the effective three-level model according to Eq.~\eqref{eq:eom}. The parameters for the simulation are shown in Appendix \ref{ap:param}. For results based on the full atom-cavity model Eq.~\eqref{eq:hamac}, we numerically determine $\epsilon_\mathrm{cr}$ from the onset of intracavity photon number \cite{Cosme2019}. Moreover, the BDW oscillator frequency $\omega_\mathrm{B}$ for the full atom-cavity model is extracted from the oscillation frequency of the BDW order parameter $\Theta_{\mathrm{BDW}}$ \cite{Cosme2019}. 
We show in Fig.~\ref{fig:npsupcav} the time-averaged occupation of the cavity mode $|a|^2$,
\begin{equation}
\widetilde{|a|^2} = \frac{1}{\tau}\int_0^\tau |a|^2\, dt,
\end{equation}
for $\tau = 100 T$, as a function of modulation strength $f_0$ and modulation frequency $\omega_{\mathrm{dr}}$.
 For $\lambda < \lambda_\mathrm{cr}$, we obtain a qualitatively similar dynamical phase diagrams for the three-level Dicke model and the full atom-cavity model, as depicted in Figs.~\ref{fig:npsupcav}(a) and \ref{fig:npsupcav}(c). Therefore, in this regime, the approximation of Eq.~\eqref{eq:hamac} via Eq.~\eqref{eq:ham2} is applicable. That is, the parametrically driven open three-level Dicke Hamiltonian is realized approximately by the shaken atom-cavity system. Moreover, the instability region from the oscillator model in the thermodynamic limit Eq.~\eqref{eq:HamHPdriven} matches the onset of the cavity mode occupation in Fig.~\ref{fig:npsupcav}(a). 

\begin{figure}[!htpb]
\centering
\includegraphics[width=1.0\columnwidth]{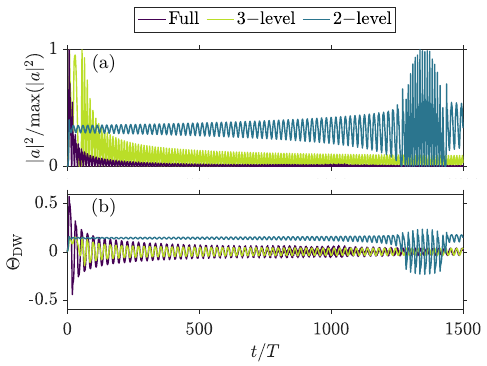}
\caption{Comparison of the dynamics between the full atom-cavity model (in purple (black)), three-level  (in green (light gray)), and coupled two-level Dicke model (in blue (dark gray)) for the (a) cavity mode occupation and (b) DW order parameter. For the Dicke models, the light-matter coupling strength is $\lambda =1.02 \lambda_{\mathrm{cr}}$. This corresponds to a pump strength of $\epsilon_{\mathrm{p}}/\epsilon_{\mathrm{cr}}=1.04$ in the full-atom cavity model. The driving parameters are fixed to $f_0=0.03$ and $\omega_{\mathrm{dr}} = 1.05 \omega_{\mathrm{B}}$.}
\label{fig:comp} 
\end{figure} 

For $\lambda > \lambda_\mathrm{cr}$, the dark areas in Figs.~\ref{fig:npsupcav}(b) and \ref{fig:npsupcav}(d) signify that the system has entered the DW  or SR phase indicated by a nonvanishing cavity mode occupation, as expected for weak and off-resonant driving. However, the DW phase is suppressed for driving frequencies blue-detuned from $\omega_{\mathrm{B}}$ as indicated  by the relative decrease in the cavity photon number in the light areas in Figs.~\ref{fig:npsupcav}(b) and \ref{fig:npsupcav}(d). Crucially, the correspondence between Eqs.~\eqref{eq:hamac} and \eqref{eq:ham2} breaks down for driving frequencies far-detuned from $\omega_\mathrm{B}$ as inferred from the parameter region $\omega_\mathrm{dr} > \omega_\mathrm{B}$ in Figs.~\ref{fig:npsupcav}(b) and \ref{fig:npsupcav}(d). This can be attributed to the occupation of higher momentum modes, specifically $|\pm 2k, 0\rangle$, in the full atom-cavity system \cite{Cosme2019}, which is not captured in the low-momentum expansion Eq.~\eqref{eq:expa} utilised in the mapping. Nevertheless, we still find good agreement on the qualitative features for driving frequencies near $\omega_\mathrm{B}$.

We also consider the dynamics according to a coupled two-level Dicke model for the same set of parameters (see Appendix \ref{ap:2comp} for details). In Fig.~\ref{fig:comp}, we present the dynamics for $\lambda > \lambda_\mathrm{cr}$ with a driving frequency blue-detuned with respect to $\omega_\mathrm{B}$. 
The results of the coupled two-level systems clearly do not capture the dynamics of the full atom-cavity system. 
On the other hand, the three-level Dicke model and the full atom-cavity model predict the same dynamical response, which is a subharmonic motion exhibited as a pulsating photon number (see Fig.~\ref{fig:comp}(a)) and a periodic switching of the sign of the DW order parameter (see Fig.~\ref{fig:comp}(b)). 
This further supports our claim that the mapping between the three-level Dicke model and the full-atom cavity system is applicable to $\lambda > \lambda_\mathrm{cr}$ for as long as the driving frequency is close to $\omega_{\mathrm{B}}$.
Note, however, that the coupled two-level systems model and the three-level model agree with each other for off-resonant driving when $j^{\mathrm{B}}_x \approx 0$, as demonstrated in Appendix \ref{ap:2comp}.

\section{Conclusions}\label{sec:con}

In this work, we have investigated a three-level Dicke model, and derived its equilibrium phase diagram, which features a normal phase and a superradiant phase. We advanced the model to a driven-dissipative system, by including a dissipation mechanism via photon loss and a periodic driving process. For this system,{we developed the dynamical phase diagram, which shows the phases for varying driving parameters}, utilizing analytical and numerical methods. As a central result we have characterized the regime of an incommensurate time crystalline state in the phase diagram. Furthermore, we obtained light-enhanced and light-induced superradiant states, in which the equilibrium superradiant state is dynamically stabilized. As a physical system that can be naturally approximated via the three-level Dicke model, we have identified a periodically shaken atom-cavity system. While the non-shaken atom-cavity system can be approximated via the standard two-level Dicke model, the shaking induces the atoms to populate additional states that are modeled via a third state in the three-level Dicke model. We note that the LISR and LESR phases display similarities with light-induced \cite{Fausti2011} and light-enhanced superconductivity \cite{Hu2014}, for which mechanisms have been proposed that involve the excitation of auxiliary modes, such as phonons \cite{Mankowsky2014,Denny2015,Okamoto2016} and Higgs bosons \cite{Homann2021}, by means of optical pumping. 
Photoexcitation of the Higgs mode in cuprate superconductors has also been predicted to lead to an incommensurate time crystal \cite{Homann2020}.
In this work, the BDW state plays the role of such an auxiliary mode, as its excitation (or equivalently, the $|1\rangle \rightarrow |3\rangle$ in Fig.~\ref{fig:schem}(a)) can be used to dynamically control the system to induce or enhance superradiance, or to enter a genuine dynamical order, namely the incommensurate time crystalline phase. We therefore expand the dynamical control of phases in atom-cavity systems to include light-induced and light-enhanced superradiance, in addition to the previously observed light-enhanced BEC or normal phase \cite{Cosme2018,Georges2018}.

\begin{acknowledgments}
We thank G. Homann and L. Broers for useful discussions.
This work is funded by the Deutsche Forschungsgemeinschaft (DFG, German Research Foundation) – SFB-925 – project 170620586 and the Cluster of Excellence “Advanced Imaging of Matter” (EXC 2056), Project No. 390715994. J.S. acknowledges support from the German Academic
Scholarship Foundation.
\end{acknowledgments}

Note: During submission of this work, a subsequent example of the driven three-level Dicke model was presented in \cite{lin2021dissipationengineered}.

\setcounter{equation}{0}
\setcounter{table}{0}
\appendix
\section{The SU(3) algebra, Gell-Mann matrices and the Casimir charges}
\label{ap:algebra}
\renewcommand{\theequation}{A\arabic{equation}}
\begin{align}\label{eq:comm}
[\hat{J}^{12}_z,\hat{J}^{12}_\pm]&=\pm\hat{J}^{12}_\pm, \quad[\hat{J}^{12}_-,\hat{J}^{12}_+]=2\hat{J}^{12}_z+\hat{J}^{13}_z+\frac{N}{2}, \nonumber \\ \nonumber
[\hat{J}^{13}_z,\hat{J}^{13}_\pm]&=\pm\hat{J}^{13}_\pm, \quad [\hat{J}^{13}_-,\hat{J}^{13}_+]=2\hat{J}^{13}_z+\hat{J}^{12}_z+\frac{N}{2}, \\ \nonumber
[\hat{J}^{12}_{\pm}, \hat{J}^{13}_{\mp}]&=\pm\hat{J}^{23}_{\pm},\quad [\hat{J}^{23}_+,\hat{J}^{23}_-]=\hat{J}^{12}_z-\hat{J}^{13}_z, \\ \nonumber
[\hat{J}^{12}_z,\hat{J}^{23}_\pm]&=\pm\hat{J}^{23}_\pm,  \quad [\hat{J}^{13}_z,\hat{J}^{23}_\pm]=\mp\hat{J}^{23}_\pm,  \\ 
[\hat{J}^{12}_\pm,\hat{J}^{23}_\mp]&=\mp\hat{J}^{13}_\pm,  \qquad [\hat{J}^{13}_\pm,\hat{J}^{23}_\pm]=\mp\hat{J}^{12}_\pm. 
\end{align}
The remaining commutators not listed above vanish. 
Our choice of pseudospin operators for the SU(3) algebra can be mapped onto the Gell-Mann matrices \cite{georgi2018lie} via
\begin{align}
&F_1 \equiv J^{12}_x = \frac{1}{2} \lambda_1,  \nonumber \\
&F_2 \equiv J^{12}_y = \frac{1}{2} \lambda_2,  \nonumber \\
&F_3 \equiv  J^{12}_z+\frac{1}{2}J^{13}_z+N/4 = \frac{1}{2} \lambda_3,  \nonumber \\
&F_4 \equiv J^{23}_x = \frac{1}{2} \lambda_4, \nonumber \\
&F_5 \equiv J^{23}_y = \frac{1}{2} \lambda_5,  \nonumber \\
&F_6 \equiv J^{13}_x = \frac{1}{2} \lambda_6, \nonumber \\
&F_7 \equiv -J^{13}_y = \frac{1}{2} \lambda_7,  \nonumber \\
&F_8 \equiv- \frac{\sqrt{3}}{2}(J^{13}_z+N/6) = \frac{1}{2} \lambda_8. 
\end{align}
\subsection{Casimir charges}
The group SU(3) enjoys two Casimirs, which can be written in matrix form using the Gell-Mann basis as
\begin{align}
C_1 &= \sum^8_{i=1} F_i F_i \\
C_2 &=\sum^8_{j,k,l=1} d_{jkl}F_j F_k F_l
\end{align}
with
\begin{equation}
d_{jkl} = \frac{1}{4} \mathrm{Tr}\left( \{\lambda_j,\lambda_k \} \lambda_l \right).
\end{equation}
\begin{widetext}
In our chosen basis, they take the form of

\begin{align}
\langle C_1 \rangle /N &=\frac{1}{12}+\left( j^{12}_x\right)^2+\left( j^{12}_y\right)^2+\left( j^{12}_z\right)^2+\left( j^{13}_x\right)^2+\left( j^{13}_y\right)^2  +\left( j^{13}_z\right)^2+\left( j^{23}_x\right)^2+\left( j^{23}_y\right)^2+\frac{1}{2}\left(j^{12}_z+j^{13}_z+2 j^{12}_zj^{13}_z \right)  \\
\langle C_2 \rangle / N^{3/2} &=\frac{1}{72} \left\{-18 \left( j^{12}_y \right)^2 + 216  j^{23}_y ( j^{12}_y  j^{13}_x -  j^{12}_x  j^{13}_y) + \nonumber
   216 j^{23}_x ( j^{12}_x  j^{13}_x + j^{12}_y j^{13}_y) \right. \\ & \left.- (1 + 6  j^{12}_z) \left[1 + 3  j^{12}_z  + \nonumber
      18 \left( j^{13}_x \right)^2 + 18 \left( j^{13}_y \right)^2\right] - 9j^{13}_z  + 36 \left( j^{23}_x \right)^2 (1 + 3 j^{12}_z + 3j^{13}_z) \right. \\&\left.+ 
   36 \left( j^{23}_y \right)^2 (1 + 3j^{12}_z + 3 j^{13}_z)- 
   18 \left[\left( j^{12}_x \right)^2 (1 + 6j^{13}_z) + \nonumber
      j^{13}_z \left(6\left(j^{12}_y \right)^2 + j^{13}_z + 2 j^{12}_z (2 + 3 j^{12}_z + 3 j^{13}_z)\right)\right]\right\}~.
\end{align}
\end{widetext}

\section{Two-Component Dicke Model}\label{ap:2comp}
\begin{figure}[!htb]
\centering
\includegraphics[width=1.0\columnwidth]{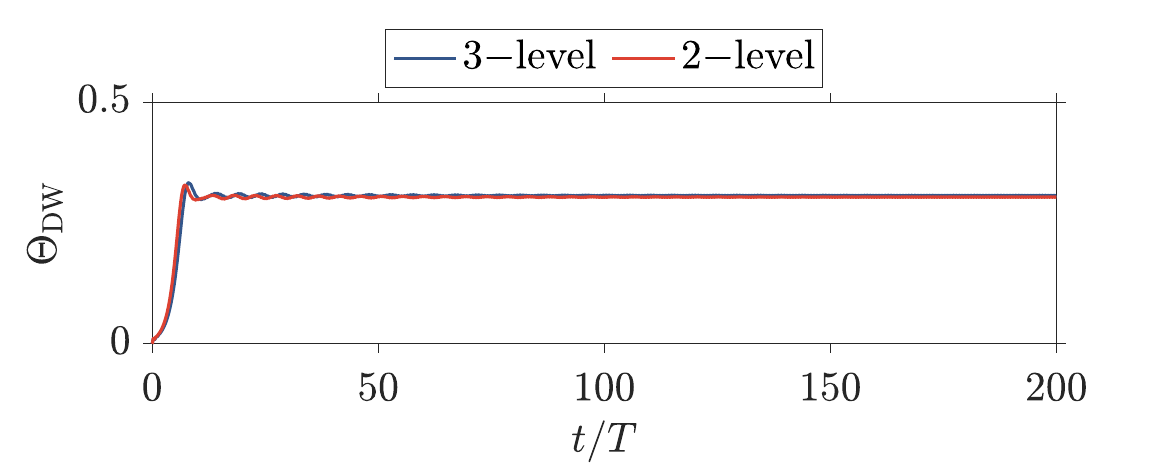}
\caption{Dynamics of the density wave order parameter for $\omega_{\mathrm{dr}} = 0.8 \omega_{\mathrm{B}}$ and $f_0=0.05$. The remaining parameters are the same as in Fig.~\ref{fig:comp}.}
\label{fig:offreso} 
\end{figure}

\renewcommand{\theequation}{B\arabic{equation}}
A modified version of the two-component Dicke model \cite{Chiacchio2019,Buca2019}, which can be realised in a spinor BEC coupled to an optical cavity \cite{Landini2018,Dogra2019}, is given by
\begin{align}\label{eq:hamsu2}
H/ \hbar &= \omega \hat{a}^\dagger \hat{a} + \omega_{\mathrm{1}} \hat{J}^{1}_z   +\omega_{\mathrm{2}}\hat{J}^{\mathrm{2}}_z  \\ \nonumber
&+\frac{2}{\sqrt{N}}\left(\hat{a}^\dagger+\hat{a} \right) \left(\lambda_{1}\hat{J}^{\mathrm{1}}_x + \lambda_{2} \hat{J}^{\mathrm{2}}_x \right).
\end{align}
Note that this has the same form as the three-level Hamiltonian in Eq.~\eqref{eq:ham} except that here, the pseudospin operators fulfil to the SU(2) group algebra with the commutation relations,
\begin{equation}\label{eq:commsu2}
[\hat{J}^{\ell}_z,\hat{J}^{\ell}_\pm]=\pm\hat{J}^{\ell}_\pm, \quad[\hat{J}^{\ell}_-,\hat{J}^{\ell}_+]=2\hat{J}^{\ell}_z,
\end{equation}
where $\ell \in \{1,2\}$.
Applying the same mean-field approximation as in Sec.~\ref{sec:eomstat}, we obtain the following equations of motion consistent with those in Refs.~\cite{Landini2018,Chiacchio2019,Buca2019,Dogra2019},
\begin{align}
\frac{d a}{dt} &= -(i\omega + \kappa)a - i2 \sum^2_{\ell=1}\lambda_\ell j^\ell_x   \\ \nonumber
\frac{d j^{\ell}_x}{dt} &= -\omega_{\mathrm{\ell}} j^{\ell}_y \\ \nonumber
\frac{d j^{\ell}_y}{dt} &= \omega_{\mathrm{\ell}} j^{\ell}_x  - 2\lambda_{\ell}(a + a^*)j^{\ell}_z \\ \nonumber
\frac{d j^{\ell}_z}{dt} &= 2\lambda_{\ell}(a + a^*)j^{\ell}_y.
\end{align}
To obtain the relevant curves in Fig.~\ref{fig:comp}, we propagate the above set of coupled equations with $\omega_1=\omega_{\mathrm{D}}$, $\omega_2=\omega_{\mathrm{B}}$, $\lambda_{1}=\lambda$, and $\lambda_2 = -\lambda f_0 \sin(\omega_{\mathrm{dr}} t)$. The exact values of these parameters are the same as those described in the main text. We present in Fig.~\ref{fig:offreso} a comparison of the dynamics according to the two-component Dicke model and the three-level Dicke model for off-resonant driving.

\section{Critical light-matter coupling}\label{ap:critlam}
\renewcommand{\theequation}{C\arabic{equation}}
Using the Hamiltonian in Eq.~(\ref{eq:HamHPSU3}) and the Heisenberg equation in  Eq.~(\ref{eq:heom}), we obtain the equations of motion as
\begin{align}
\frac{d a}{dt} &= -(i\omega -\kappa)a-i \lambda_{12} \left(a^\dagger_{12}+a_{12} \right)-i \lambda_{13} \left(a^\dagger_{13}+a_{13} \right) \\ \nonumber
\frac{d a_{12}}{dt} &=-i \omega_{12}a_{12}-i \lambda_{12} \left(a^*+a \right)\\ \nonumber
\frac{d a_{13}}{dt} &=-i \omega_{13}a_{13}-i \lambda_{13} \left(a^*+a \right)
\end{align}
We can then construct the matrix $M$ as $\partial_t \vec{v}= M \vec{v}$ to obtain
\begin{equation}
M=\left(
\begin{array}{cccccc}
\kappa-i \omega & 0 & -i \lambda_{12}  & -i \lambda_{12}  & -i \lambda_{13}  & -i \lambda_{13}  \\
 0 & \kappa+i \omega & i \lambda_{12}& i \lambda_{12}& i \lambda_{13} & i \lambda_{13} \\
 -i \lambda_{12}  & -i \lambda_{12}  & -i \omega_{12}& 0 & 0 & 0 \\
 i \lambda_{12}  & i \lambda_{12}  & 0 & i \omega_{12}& 0 & 0 \\
 -i \lambda_{13} & -i \lambda_{13} & 0 & 0 & -i \omega_{13} & 0 \\
 i \lambda_{13} & i \lambda_{13} & 0 & 0 & 0 & i \omega_{13} \\
\end{array}
\right)~.
\end{equation}
A phase transition occurs if $M$ inherits a zero energy eigenstate \cite{Kirton2019}. This means, to find the critical light-matter coupling $\lambda$, we need to calculate $\mathrm{det}(M)=0$, giving us
\begin{equation}
\frac{\left( \kappa^2+\omega^2 \right)}{4 \omega}= \left( \frac{\lambda^2_{12}}{\omega_{12}}+\frac{\lambda^2_{13}}{\omega_{13}} \right).
\end{equation}

\section{Heisenberg Equations of Motion}\label{ap:eom}
\renewcommand{\theequation}{D\arabic{equation}}
The dynamics of an observable $\hat{O}$ in the dissipative system considered here is governed by the Heisenberg equation
\begin{equation}\label{eq:heom}
\frac{d\langle \hat{O} \rangle}{dt} = \biggl\langle \frac{i}{\hbar}[\hat{H},\hat{O}] + \kappa\left( 2\hat{a}^{\dagger} \hat{O} \hat{a} - \{\hat{a}^{\dagger} \hat{a},\hat{O} \} \right) \biggr\rangle.
\end{equation}
Using the commutation relations Eq.~\eqref{eq:comm}, we get the following equations for the expectation values of relevant operators in the open three-level Dicke model Eq.~\eqref{eq:ham}
\begin{widetext}
\begin{align}
\frac{d \langle a \rangle}{dt}&= -(i \omega+\kappa) \langle a \rangle -i \frac{2 }{\sqrt{N}}\biggl(\lambda_{12} \langle \hat{J}^{12}_x \rangle +  \lambda_{13}\langle \hat{J}^{13}_x \rangle  \biggr)
\\ \nonumber
\frac{d \langle \hat{J}^{12}_x \rangle}{dt}&= -\omega_{12}\langle \hat{J}^{12}_y \rangle+ \frac{\lambda_{13}}{\sqrt{N}}\langle (a^\dagger+a)\hat{J}^y_{23}\rangle
\\ \nonumber
\frac{d \langle \hat{J}^{12}_y \rangle}{dt}&=\omega_{12}\langle \hat{J}^{12}_x \rangle-\frac{\lambda_{12}}{\sqrt{N}}\left(2\langle (a^\dagger+a)\hat{J}^{12}_z \rangle+\langle (a^\dagger+a)\hat{J}^{13}_z \rangle+\langle (a^\dagger+a)N/2\rangle \right)-\frac{\lambda_{13}}{\sqrt{N}}\langle (a^\dagger+a)\hat{J}^{23}_{x}\rangle
\\ \nonumber
\frac{d \langle \hat{J}^{12}_z \rangle}{dt}&= \frac{2 \lambda_{12}}{\sqrt{N}} \langle (a+a^\dagger)\hat{J}^{12}_y \rangle
\\  \nonumber
\frac{d \langle \hat{J}^{13}_x \rangle}{dt}&= -\omega_{13}\langle \hat{J}^{13}_y \rangle- \frac{\lambda_{12}}{\sqrt{N}}\langle (a^\dagger+a)\hat{J}^{23}_{y}\rangle
\\ \nonumber
\frac{d \langle \hat{J}^{13}_y \rangle}{dt}&=\omega_{13}\langle \hat{J}^{13}_x \rangle-\frac{\lambda_{13}}{\sqrt{N}}\left(2\langle (a^\dagger+a)\hat{J}^{13}_z \rangle+\langle (a^\dagger+a)\hat{J}^{12}_z \rangle+\langle (a^\dagger+a)N/2\rangle \right)   -\frac{\lambda_{12}}{\sqrt{N}}\langle (a^\dagger+a)\hat{J}^{23}_{x}\rangle
\\ \nonumber
\frac{d \langle \hat{J}^{13}_z \rangle}{dt}&= \frac{2 \lambda_{13}}{\sqrt{N}} \langle (a+a^\dagger)\hat{J}^{13}_y \rangle
\\ \nonumber
\frac{d \langle \hat{J}^{23}_x \rangle}{dt}&= (\omega_{13}-\omega_{12}) \langle \hat{J}^{23}_y \rangle +\frac{\lambda_{12}}{\sqrt{N}}\langle (a^\dagger+a) \hat{J}^{13}_y \rangle+\frac{\lambda_{13}}{\sqrt{N}}\langle (a^\dagger+a) \hat{J}^{12}_y \rangle
\\ \nonumber
\frac{d \langle \hat{J}^{23}_y \rangle}{dt}&=  (\omega_{12}-\omega_{13})\langle \hat{J}^{23}_y \rangle +\frac{\lambda_{12}}{\sqrt{N}}\langle (a^\dagger+a) \hat{J}^{13}_x \rangle- \frac{\lambda_{13}}{\sqrt{N}}\langle (a^\dagger+a) \hat{J}^{12}_x \rangle 
\end{align}
On the other hand, the equations of motion for the parametrically driven open three-level Dicke model are
\begin{align}
\frac{d \langle \hat{a} \rangle}{dt}&= -(i \omega +\kappa)\langle \hat{a} \rangle-i \frac{2 \lambda}{\sqrt{N}} \langle \hat{J}^{\mathrm{D}}_x \rangle+i \phi(t) \frac{2 \lambda}{\sqrt{N}} \langle \hat{J}^{\mathrm{B}}_x \rangle \\ \nonumber
\frac{d \langle \hat{J}^\mathrm{D}_x \rangle}{dt}&= -\omega_\mathrm{D}\langle \hat{J}^\mathrm{D}_y \rangle-\phi(t)(\omega_{\mathrm{B}}-\omega_{\mathrm{D}}) \langle \hat{J}^\mathrm{B}_y \rangle -\phi(t) \frac{\lambda}{\sqrt{N}}\langle (a^\dagger+a)\hat{J}^{\mathrm{DB}}_y\rangle
\\ \nonumber
\frac{d \langle \hat{J}^\mathrm{D}_y \rangle}{dt}&=\omega_\mathrm{D}\langle \hat{J}^\mathrm{D}_x \rangle-\frac{\lambda}{\sqrt{N}}\biggl[2\langle (a^\dagger+a)\hat{J}^\mathrm{D}_z \rangle +\langle (a^\dagger+a)\hat{J}^\mathrm{B}_z \rangle+\langle a^\dagger+a\rangle\frac{N}{2} \biggr] +\phi(t)(\omega_{\mathrm{B}}-\omega_{\mathrm{D}}) \langle \hat{J}^\mathrm{B}_x \rangle+\phi(t) \frac{\lambda}{\sqrt{N}}\langle (a^\dagger+a)\hat{J}^{\mathrm{DB}}_x\rangle
\\  \nonumber
\frac{d \langle \hat{J}^\mathrm{D}_z \rangle}{dt}&= \frac{2 \lambda}{\sqrt{N}} \langle (a+a^\dagger)\hat{J}^\mathrm{D}_y \rangle+ 2 (\omega_{\mathrm{B}}-\omega_{\mathrm{D}})\phi(t) \langle \hat{J}^{\mathrm{DB}}_y \rangle 
\\  \nonumber
\frac{d \langle \hat{J}^\mathrm{B}_x \rangle}{dt}&= -\omega_\mathrm{B}\langle \hat{J}^\mathrm{B}_y \rangle-\phi(t) (\omega_{\mathrm{B}}-\omega_{\mathrm{D}})\langle \hat{J}^\mathrm{D}_y \rangle - \frac{\lambda}{\sqrt{N}}\langle (a^\dagger+a)\hat{J}^{\mathrm{DB}}_y\rangle
\\ \nonumber
\frac{d \langle \hat{J}^\mathrm{B}_y \rangle}{dt}&=\omega_\mathrm{B}\langle \hat{J}^\mathrm{B}_x \rangle+\phi(t)\frac{\lambda}{\sqrt{N}}\biggl[2\langle (a^\dagger+a)\hat{J}^\mathrm{B}_z \rangle+\langle (a^\dagger+a)\hat{J}^\mathrm{D}_z \rangle+\langle a^\dagger+a\rangle\frac{N}{2} \biggr]  +\phi(t) (\omega_{\mathrm{B}}-\omega_{\mathrm{D}})\langle \hat{J}^\mathrm{D}_x \rangle -\frac{\lambda}{\sqrt{N}}\langle (a^\dagger+a)\hat{J}^{\mathrm{DB}}_x\rangle
\\ \nonumber
\frac{d \langle \hat{J}^\mathrm{B}_z \rangle}{dt}&= -\phi(t)  (\omega_{\mathrm{B}}-\omega_{\mathrm{D}}) \langle (a+a^\dagger)\hat{J}^\mathrm{B}_y \rangle- \frac{4 \lambda^2}{U_0 N}\phi(t) \langle \hat{J}^{\mathrm{DB}}_y \rangle
\\ \nonumber
\frac{d \langle \hat{J}^\mathrm{DB}_x \rangle}{dt}&= (\omega_{\mathrm{B}}-\omega_{\mathrm{D}}) \langle \hat{J}^\mathrm{DB}_y \rangle +\frac{\lambda}{\sqrt{N}}\langle (a^\dagger+a) \hat{J}^\mathrm{B}_y \rangle -\phi(t) \frac{\lambda}{\sqrt{N}}\langle (a^\dagger+a) \hat{J}^\mathrm{D}_y \rangle
\\ \nonumber
\frac{d \langle \hat{J}^\mathrm{DB}_y \rangle}{dt}&= (\omega_{\mathrm{D}}-\omega_{\mathrm{B}})\langle \hat{J}^\mathrm{DB}_x \rangle +\frac{\lambda}{\sqrt{N}}\langle (a^\dagger+a) \hat{J}^\mathrm{B}_x \rangle +\phi(t) \frac{\lambda}{\sqrt{N}}\langle (a^\dagger+a) \hat{J}^\mathrm{D}_x \rangle + 2 (\omega_{\mathrm{B}}-\omega_{\mathrm{D}}) \phi(t) \langle \hat{J}^\mathrm{B}_z-\hat{J}^\mathrm{D}_z \rangle
\end{align}
\end{widetext}

\section{Mean-field equations of motion}
\label{ap:mf_eom}
\renewcommand{\theequation}{E\arabic{equation}}
The mean-field equations for the dissipative three-level Dicke model are given by
\begin{align}\label{eq:eom1}
\frac{d a }{dt}&= -(i \omega+\kappa)a -i 2 \lambda_{12}  j^{12}_x -i  2 \lambda_{13}  j^{13}_x \\ \nonumber
\frac{d  j^{12}_x }{dt}&= -\omega_{12} j^{12}_y + \lambda_{13} (a+a^*)j^y_{23}
\\ \nonumber
\frac{d j^{12}_y }{dt}&=\omega_{12} j^{12}_x -\lambda_{12} (a+a^*)\left(2 j^{12}_z +j^{13}_z +1/2 \right) \\ \nonumber
&\qquad-\lambda_{13} (a+a^*)j^{23}_{x}
\\ \nonumber
\frac{d j^{12}_z }{dt}&= 2 \lambda_{12}  (a+a^*)j^{12}_y 
\\ \nonumber
\frac{d  j^{13}_x }{dt}&= -\omega_{13} j^{13}_y  - \lambda_{12}  (a+a^*)j^{23}_{y}
\\ \nonumber
\frac{d  j^{13}_y }{dt}&=\omega_{13} j^{13}_x -\lambda_{13} (a+a^*) \left(2 j^{13}_z + j^{12}_z + 1/2 \right) \\ \nonumber
&\qquad  -\lambda_{12}(a+a^*)j^{23}_{x}
\\ \nonumber
\frac{d  j^{13}_z }{dt}&= 2 \lambda_{13}  (a+a^*)j^{13}_y 
\\ \nonumber
\frac{d  j^{23}_x }{dt}&= (\omega_{13}-\omega_{12})  j^{23}_y  +\lambda_{12}  (a+a^*) j^{13}_y +\lambda_{13} (a+a^*) j^{12}_y 
\\ \nonumber
\frac{d  j^{23}_y }{dt}&=  (\omega_{12}-\omega_{13}) j^{23}_y  +\lambda_{12}  (a+a^*)j^{13}_x - \lambda_{13} (a+a^*) j^{12}_x 
\end{align}

For the parametrically driven open three-level Dicke model, the equations of motion are given by
\begin{align}\label{eq:eom}
\frac{d a }{dt}&= -(i \omega +\kappa) a -i 2 \lambda j^\mathrm{D}_x +i \phi(t) 2 \lambda  j^\mathrm{B}_x  
\\ \nonumber
\frac{d  j^\mathrm{D}_x }{dt}&= -\omega_\mathrm{D} j^\mathrm{D}_y - \phi(t) (\omega_{\mathrm{B}}-\omega_{\mathrm{D}}) j^\mathrm{B}_y -\phi(t) \lambda (a+a^*)j^{\mathrm{DB}}_y
\\ \nonumber
\frac{d  j^\mathrm{D}_y }{dt}&=\omega_\mathrm{D} j^\mathrm{D}_x-\lambda(a+a^*)\biggl(2 j^\mathrm{D}_z + j^\mathrm{B}_z +1/2 \biggr) 
\\ \nonumber
&\qquad +\phi(t) (\omega_{\mathrm{B}}-\omega_{\mathrm{D}})  j^\mathrm{B}_x +\phi(t)\lambda (a+a^*)j^\mathrm{DB}_x
\\ \nonumber
\frac{d j^\mathrm{D}_z }{dt}&= 2 \lambda  (a+a^*)j^\mathrm{D}_y + 2 (\omega_{\mathrm{B}}-\omega_{\mathrm{D}})\phi(t)  j^{\mathrm{DB}}_y 
\\ \nonumber
\frac{d j^\mathrm{B}_x }{dt}&= -\omega_\mathrm{B} j^\mathrm{B}_y -(\omega_{\mathrm{B}}-\omega_{\mathrm{D}})\phi(t) j^\mathrm{D}_y - \lambda (a+a^*)j^{\mathrm{DB}}_y
\\ \nonumber
\frac{d  j^\mathrm{B}_y }{dt}&=\omega_\mathrm{B} j^\mathrm{B}_x +\phi(t)\lambda (a+a^*)\biggl(2j^\mathrm{B}_z + j^\mathrm{D}_z +1/2 \biggr)  
\\ \nonumber
&\qquad +\phi(t) (\omega_{\mathrm{B}}-\omega_{\mathrm{D}})  j^\mathrm{D}_x  -\lambda(a+a^*) j^\mathrm{DB}_x
\\ \nonumber
\frac{d  j^\mathrm{B}_z}{dt}&= -\phi(t) 2 \lambda (a+a^*)j^\mathrm{B}_y - 2 (\omega_{\mathrm{B}}-\omega_{\mathrm{D}})\phi(t)  j^{\mathrm{DB}}_y 
\\ \nonumber
\frac{d j^{\mathrm{DB}}_x }{dt}&=  (\omega_{\mathrm{B}}-\omega_{\mathrm{D}}) j^{\mathrm{DB}}_y  +\lambda(a+a^*) j^\mathrm{B}_y -\phi(t) \lambda (a+a^*) j^\mathrm{D}_y
\\ \nonumber
\frac{d j^{\mathrm{DB}}_y }{dt}&= (\omega_{\mathrm{D}}-\omega_{\mathrm{B}})  j^{\mathrm{DB}}_x  +\lambda(a+a^*) j^\mathrm{B}_x +\phi(t) \lambda(a+a^*) j^\mathrm{D}_x 
\\ \nonumber
& \qquad + 2 (\omega_{\mathrm{B}}-\omega_{\mathrm{D}}) \phi(t)  (j^\mathrm{B}_z-j^\mathrm{D}_z).
\end{align}

\begin{figure}[!htb]
\centering
\includegraphics[width=1.0\columnwidth]{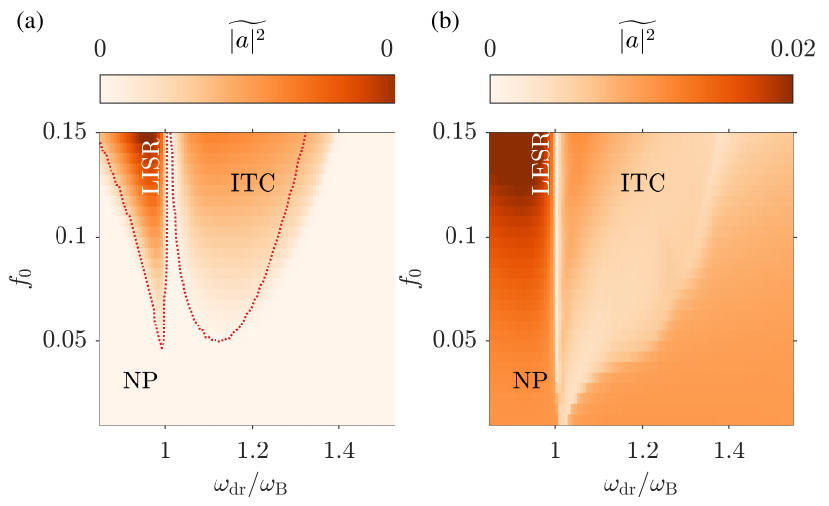}
\caption{Time-averaged cavity mode occupation $|a|^2$ taken over 100 driving cycles, $\tau = 100T$. We choose $\omega_\mathrm{D}= \omega_\mathrm{B}=\omega_{\mathrm{rec}}$ while the remaining parameters are the same as those in Figs.~\ref{fig:npsupcav}(a)-(b).}
\label{fig:w_equal} 
\end{figure}

{In Fig.~\ref{fig:w_equal}, we demonstrate that the existence of the dynamical phases is independent of the term in the Hamiltonian with $j^{\mathrm{DB}}_x$. That is, the momenta coupling inferred from Eq.~\eqref{eq:hamhpxp} does not play a crucial role in the formation of the ITC, LESR, and LISR phases. This suggests that the emergence of these dynamical phases originates from the last term in Eq.~\eqref{eq:ham2}.
To confirm this, we set $\omega_\mathrm{D}= \omega_\mathrm{B}$ in Fig.~\ref{fig:w_equal}. For comparison, we show in dotted lines the phase boundary obtained for $\omega_\mathrm{D} \neq \omega_\mathrm{B}$.}

\section{Holstein-Primakoff Transformation} \label{ap:hp}
\renewcommand{\theequation}{F\arabic{equation}}

We present a Holstein-Primakoff approximation in the thermodynamic limit, i.e. $N \rightarrow \infty$ \cite{Emary2003,Bastidas2012}. To capture the correct SU(3) algebra, we use an extended version of the Holstein-Primakoff representation given by \cite{Wagner1975}
\begin{align}
\hat{J}^{12}_z &= \hat{a}^\dagger_{12} \hat{a}_{12}-N/2, \quad \hat{J}^{12}_+ = \hat{a}^\dagger_{12} ~\sqrt{N-\left( \hat{a}^\dagger_{12}  \hat{a}_{12}+\hat{a}^\dagger_{13}  \hat{a}_{13} \right)} \\ \nonumber
\hat{J}^{12}_- &=  \sqrt{N-\left( \hat{a}^\dagger_{12}  \hat{a}_{12}+\hat{a}^\dagger_{13} \hat{a}_{13} \right)}~\hat{a}_{12}\\ \nonumber
\hat{J}^{13}_z &= \hat{a}^\dagger_{13} \hat{a}_{13}-N/2, \quad \hat{J}^{13}_+ = \hat{a}^\dagger_{13}~\sqrt{N-\left( \hat{a}^\dagger_{12} \hat{a}_{12}+\hat{a}^\dagger_{13}  \hat{a}_{13} \right)} \\ \nonumber
\hat{J}^{13}_- &=  \sqrt{N-\left( \hat{a}^\dagger_{12} \hat{a}_{12}+\hat{a}^\dagger_{13}  \hat{a}_{13}\right)}~  \hat{a}_{13} \\ \nonumber
\hat{J}^{23}_+ &= \hat{a}^\dagger_{12}   \hat{a}_{13}, \quad \hat{J}^{23}_- = \hat{a}^\dagger_{13}\hat{a}_{12}.
\end{align}
In the $N \to \infty$ limit, we can further approximate the pseudospin operators as
\begin{align}
\hat{J}^{\mathrm{D}}_z &=\hat{a}^\dagger_{12}  \hat{a}_{12} -N/2, \quad \hat{J}^{\mathrm{D}}_+ = \hat{a}^\dagger_{12}  ~\sqrt{N}, \quad  \hat{J}^{\mathrm{D}}_- =  \sqrt{N}~\hat{a}_{12} \\ \nonumber
\hat{J}^{\mathrm{B}}_z &= \hat{a}^\dagger_{13}  \hat{a}_{13} -N/2, \quad \hat{J}^{\mathrm{B}}_+ = \hat{a}^\dagger_{13} ~\sqrt{N}, \quad \hat{J}^{\mathrm{B}}_- =  \sqrt{N}~ \hat{a}_{13} \\ \nonumber
\hat{J}^{\mathrm{DB}}_+ &=  \hat{a}^\dagger_{12}  \hat{a}_{13}, \qquad \hat{J}^{\mathrm{DB}}_- = \hat{a}^\dagger_{13} \hat{a}_{12} .
\end{align}
In an analogue fashion for the driven three-level Dicke model we obtain the Hamiltonian {with $a_{12} \rightarrow d$ and $a_{13} \rightarrow b$}
\begin{align}
\label{eq:HamHPdriven}
H/ \hbar &= \omega \hat{a}^\dagger \hat{a}+\omega_\mathrm{D} \hat{d}^\dagger\hat{d}+\omega_\mathrm{B} \hat{b}^\dagger \hat{b} +\lambda \left(\hat{a}^\dagger+ \hat{a} \right) \\ \nonumber 
&\qquad \times \left[  (\hat{d}^\dagger+\hat{d})-\phi(t)(\hat{b}^\dagger+\hat{b}) \right] \\ \nonumber
&\qquad +\phi(t)(\omega_{\mathrm{B}}-\omega_{\mathrm{D}}) \left( \hat{d}^\dagger\hat{b}+\hat{b}^\dagger\hat{d}\right) ~.
\end{align}
The mean-field equations of motion for Eq.~\eqref{eq:HamHPdriven} are
\begin{align}\label{eq:eomhpdriven}
\frac{\partial a}{\partial t} &= -(i\omega  -\kappa )a -i \lambda(d^* + d) +i \phi(t)  \lambda({b}^{*}+{b}) \\  \nonumber
\frac{\partial d}{\partial t} &=  -i\omega_{\mathrm{D}} d -i \lambda(a^* + a) -i(\omega_\mathrm{B}-\omega_{\mathrm{D}})  \phi(t)b. \\ \nonumber
\frac{\partial b}{\partial t} &= -i\omega_{\mathrm{B}} b +i  \phi(t)  \lambda ({a}^{*}+{a})-i(\omega_\mathrm{B}-\omega_{\mathrm{D}}) \phi(t) d.
\end{align}

\section{Lower Polariton}\label{ap:lp}
\renewcommand{\theequation}{G\arabic{equation}}

Consider the standard closed Dicke model
\begin{equation}
\hat{H}/\hbar= \omega \hat{a}^{\dagger}\hat{a}+ \omega_{0} \hat{J}_z  +\frac{2\lambda}{\sqrt{N}}(\hat{a}^{\dagger}+\hat{a})(\hat{J}_x).
\end{equation}
\begin{figure}[!htb]
\centering
\includegraphics[width=1.0\columnwidth]{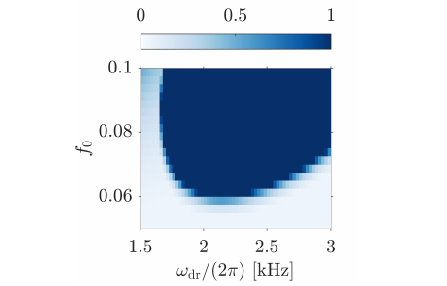}
\caption{Maximum value of $(b+b^*)$. The parameters are the same as those in Figs.~\ref{fig:norm}(a)-(d).}
\label{fig:lp} 
\end{figure} 
In the thermodynamic limit, this can be diagonalised using the Holstein-Primakoff transformation, which leads to two polariton frequencies
\begin{equation}\label{eq:lp}
\omega_{\mathrm{LP},\kappa=0} = \biggl(\frac{(\omega^2_0+\omega^2)}{2} - \frac{1}{2}\sqrt{(\omega^2_0-\omega^2)^2+16\lambda^2\omega\omega_0}\biggr)^{1/2}
\end{equation}
\begin{equation}
\omega_{\mathrm{UP},\kappa=0} = \biggl(\frac{(\omega^2_0+\omega^2)}{2} + \frac{1}{2}\sqrt{(\omega^2_0-\omega^2)^2+16\lambda^2\omega\omega_0}\biggr)^{1/2}.
\end{equation}

The lower polariton frequency, Eq.~\eqref{eq:lp}, is the upper bound in the presence of dissipation. When $\kappa \neq 0$, the lower polariton frequency can be numerically determined by exploiting the parametric resonance when the light-matter coupling is periodically driven \cite{Bastidas2012,Chitra2015}
\begin{equation}
\lambda(t) =  \lambda_0 (1+f_0 \sin(\omega_{dr}t)).
\end{equation}
In the limit $N \to \infty$, the Hamiltonian can be reduced to a coupled oscillator, whereby the coupling strength is periodic in time. This possesses a parametric resonance manifesting as a resonance lobe centred at the primary resonance, $\omega_{\mathrm{dr}} = 2\omega_\mathrm{LP}$. Thus, we can determine $\omega_\mathrm{LP}$ by mapping the instability region for varying modulation parameters $f_0$ and $\omega_{\mathrm{dr}}$. To this end, we solve the corresponding equations of motion given by
\begin{align}
i\frac{\partial a}{\partial t} & = (\omega -i\kappa)a + \lambda(t)(b^* + b) \\ \nonumber
i\frac{\partial b}{\partial t} &= \omega_0 b + \lambda(t)(a^* + a)
\end{align}
The unstable region indicating the parametric resonance is signalled by a diverging $(b+b^*)$, as depicted in Fig.~\ref{fig:lp}. We obtain a lower polariton frequency $\omega_{\mathrm{LP}}/2\pi \approx 1.06~\mathrm{kHz}$, which is the value used in the sum frequency condition denoted by the vertical line in Figs.~\ref{fig:norm}(a)-(d).

\section{Parameters}\label{ap:param}
\renewcommand{\theequation}{H\arabic{equation}}

We consider realistic parameters based on the experimental setup in Ref.~\cite{joint}. A BEC of $N = 65\times 10^3$  $^{87}$Rb atoms is coupled to a high-finesse optical cavity with a photon loss rate of $\kappa=~2\pi \times 3.6~\mathrm{kHz}$. This is very close to the recoil frequency, $\omega_\mathrm{rec} =2\pi \times 3.55~\mathrm{kHz}$, associated with the standing wave potential of the pump. The cavity light shift per atom is $U_0 = -2\pi \times 0.36~\mathrm{Hz}$. The effective pump-cavity detuning is fixed to $\delta_{\mathrm{eff}} =-2 \pi \times 18.5~\mathrm{kHz}$. We are interested in the two regimes $\lambda < \lambda_{\mathrm{cr}}$ and  $\lambda > \lambda_{\mathrm{cr}}$, where $\lambda_\mathrm{cr}$ is the critical light-matter coupling strength needed to enter the DW phase in the absence of modulation, where $\lambda_\mathrm{cr}= \sqrt{(\kappa^2+\omega^2)\left(\omega_\mathrm{D}/\omega\right)}/2$. By equating the expression for $\lambda_\mathrm{cr}$ and $\lambda$ in terms of the atom-cavity parameters for the two-level Dicke model, we find that the critical pump strength is given by  $\epsilon_{\mathrm{cr}}= {8(\omega^2+\kappa^2) }/{\left(4N\omega |\Delta_0|+ {(\omega^2+\kappa^2) }\right)} $. 

\bibliography{biblio}

\providecommand{\noopsort}[1]{}\providecommand{\singleletter}[1]{#1}%
\begin{thebibliography}{57}%
\makeatletter
\providecommand \@ifxundefined [1]{%
 \@ifx{#1\undefined}
}%
\providecommand \@ifnum [1]{%
 \ifnum #1\expandafter \@firstoftwo
 \else \expandafter \@secondoftwo
 \fi
}%
\providecommand \@ifx [1]{%
 \ifx #1\expandafter \@firstoftwo
 \else \expandafter \@secondoftwo
 \fi
}%
\providecommand \natexlab [1]{#1}%
\providecommand \enquote  [1]{``#1''}%
\providecommand \bibnamefont  [1]{#1}%
\providecommand \bibfnamefont [1]{#1}%
\providecommand \citenamefont [1]{#1}%
\providecommand \href@noop [0]{\@secondoftwo}%
\providecommand \href [0]{\begingroup \@sanitize@url \@href}%
\providecommand \@href[1]{\@@startlink{#1}\@@href}%
\providecommand \@@href[1]{\endgroup#1\@@endlink}%
\providecommand \@sanitize@url [0]{\catcode `\\12\catcode `\$12\catcode
  `\&12\catcode `\#12\catcode `\^12\catcode `\_12\catcode `\%12\relax}%
\providecommand \@@startlink[1]{}%
\providecommand \@@endlink[0]{}%
\providecommand \url  [0]{\begingroup\@sanitize@url \@url }%
\providecommand \@url [1]{\endgroup\@href {#1}{\urlprefix }}%
\providecommand \urlprefix  [0]{URL }%
\providecommand \Eprint [0]{\href }%
\providecommand \doibase [0]{http://dx.doi.org/}%
\providecommand \selectlanguage [0]{\@gobble}%
\providecommand \bibinfo  [0]{\@secondoftwo}%
\providecommand \bibfield  [0]{\@secondoftwo}%
\providecommand \translation [1]{[#1]}%
\providecommand \BibitemOpen [0]{}%
\providecommand \bibitemStop [0]{}%
\providecommand \bibitemNoStop [0]{.\EOS\space}%
\providecommand \EOS [0]{\spacefactor3000\relax}%
\providecommand \BibitemShut  [1]{\csname bibitem#1\endcsname}%
\let\auto@bib@innerbib\@empty
\bibitem [{\citenamefont {Dicke}(1954)}]{Dicke}%
  \BibitemOpen
  \bibfield  {author} {\bibinfo {author} {\bibfnamefont {R.~H.}\ \bibnamefont
  {Dicke}},\ }\bibfield  {title} {\enquote {\bibinfo {title} {Coherence in
  spontaneous radiation processes},}\ }\href {\doibase 10.1103/PhysRev.93.99}
  {\bibfield  {journal} {\bibinfo  {journal} {Phys. Rev.}\ }\textbf {\bibinfo
  {volume} {93}},\ \bibinfo {pages} {99--110} (\bibinfo {year}
  {1954})}\BibitemShut {NoStop}%
\bibitem [{\citenamefont {{Baumann}}\ \emph {et~al.}(2010)\citenamefont
  {{Baumann}}, \citenamefont {{Guerlin}}, \citenamefont {{Brennecke}},\ and\
  \citenamefont {{Esslinger}}}]{Baumann2010}%
  \BibitemOpen
  \bibfield  {author} {\bibinfo {author} {\bibfnamefont {K.}~\bibnamefont
  {{Baumann}}}, \bibinfo {author} {\bibfnamefont {C.}~\bibnamefont
  {{Guerlin}}}, \bibinfo {author} {\bibfnamefont {F.}~\bibnamefont
  {{Brennecke}}}, \ and\ \bibinfo {author} {\bibfnamefont {T.}~\bibnamefont
  {{Esslinger}}},\ }\bibfield  {title} {\enquote {\bibinfo {title} {{Dicke
  quantum phase transition with a superfluid gas in an optical cavity}},}\
  }\href {\doibase 10.1038/nature09009} {\bibfield  {journal} {\bibinfo
  {journal} {Nature}\ }\textbf {\bibinfo {volume} {464}},\ \bibinfo {pages}
  {1301--1306} (\bibinfo {year} {2010})}\BibitemShut {NoStop}%
\bibitem [{\citenamefont {{Klinder}}\ \emph {et~al.}(2015)\citenamefont
  {{Klinder}}, \citenamefont {{Ke{\ss}ler}}, \citenamefont {{Wolke}},
  \citenamefont {{Mathey}},\ and\ \citenamefont {{Hemmerich}}}]{Klinder2015}%
  \BibitemOpen
  \bibfield  {author} {\bibinfo {author} {\bibfnamefont {J.}~\bibnamefont
  {{Klinder}}}, \bibinfo {author} {\bibfnamefont {H.}~\bibnamefont
  {{Ke{\ss}ler}}}, \bibinfo {author} {\bibfnamefont {M.}~\bibnamefont
  {{Wolke}}}, \bibinfo {author} {\bibfnamefont {L.}~\bibnamefont {{Mathey}}}, \
  and\ \bibinfo {author} {\bibfnamefont {A.}~\bibnamefont {{Hemmerich}}},\
  }\bibfield  {title} {\enquote {\bibinfo {title} {{Dynamical phase transition
  in the open Dicke model}},}\ }\href {\doibase 10.1073/pnas.1417132112}
  {\bibfield  {journal} {\bibinfo  {journal} {Proc. Natl. Acad. Sci. USA}\
  }\textbf {\bibinfo {volume} {112}},\ \bibinfo {pages} {3290--3295} (\bibinfo
  {year} {2015})}\BibitemShut {NoStop}%
\bibitem [{\citenamefont {Hayn}\ \emph {et~al.}(2011)\citenamefont {Hayn},
  \citenamefont {Emary},\ and\ \citenamefont {Brandes}}]{Hayn2011}%
  \BibitemOpen
  \bibfield  {author} {\bibinfo {author} {\bibfnamefont {M.}~\bibnamefont
  {Hayn}}, \bibinfo {author} {\bibfnamefont {C.}~\bibnamefont {Emary}}, \ and\
  \bibinfo {author} {\bibfnamefont {T.}~\bibnamefont {Brandes}},\ }\bibfield
  {title} {\enquote {\bibinfo {title} {Phase transitions and dark-state physics
  in two-color superradiance},}\ }\href {\doibase 10.1103/PhysRevA.84.053856}
  {\bibfield  {journal} {\bibinfo  {journal} {Phys. Rev. A}\ }\textbf {\bibinfo
  {volume} {84}},\ \bibinfo {pages} {053856} (\bibinfo {year}
  {2011})}\BibitemShut {NoStop}%
\bibitem [{\citenamefont {Bastidas}\ \emph {et~al.}(2012)\citenamefont
  {Bastidas}, \citenamefont {Emary}, \citenamefont {Regler},\ and\
  \citenamefont {Brandes}}]{Bastidas2012}%
  \BibitemOpen
  \bibfield  {author} {\bibinfo {author} {\bibfnamefont {V.~M.}\ \bibnamefont
  {Bastidas}}, \bibinfo {author} {\bibfnamefont {C.}~\bibnamefont {Emary}},
  \bibinfo {author} {\bibfnamefont {B.}~\bibnamefont {Regler}}, \ and\ \bibinfo
  {author} {\bibfnamefont {T.}~\bibnamefont {Brandes}},\ }\bibfield  {title}
  {\enquote {\bibinfo {title} {Nonequilibrium quantum phase transitions in the
  dicke model},}\ }\href {\doibase 10.1103/PhysRevLett.108.043003} {\bibfield
  {journal} {\bibinfo  {journal} {Phys. Rev. Lett.}\ }\textbf {\bibinfo
  {volume} {108}},\ \bibinfo {pages} {043003} (\bibinfo {year}
  {2012})}\BibitemShut {NoStop}%
\bibitem [{\citenamefont {Chitra}\ and\ \citenamefont
  {Zilberberg}(2015)}]{Chitra2015}%
  \BibitemOpen
  \bibfield  {author} {\bibinfo {author} {\bibfnamefont {R.}~\bibnamefont
  {Chitra}}\ and\ \bibinfo {author} {\bibfnamefont {O.}~\bibnamefont
  {Zilberberg}},\ }\bibfield  {title} {\enquote {\bibinfo {title} {{Dynamical
  many-body phases of the parametrically driven, dissipative Dicke model}},}\
  }\href {\doibase 10.1103/PhysRevA.92.023815} {\bibfield  {journal} {\bibinfo
  {journal} {Phys. Rev. A}\ }\textbf {\bibinfo {volume} {92}},\ \bibinfo
  {pages} {023815} (\bibinfo {year} {2015})}\BibitemShut {NoStop}%
\bibitem [{\citenamefont {{Zhiqiang}}\ \emph {et~al.}(2017)\citenamefont
  {{Zhiqiang}}, \citenamefont {{Lee}}, \citenamefont {{Kumar}}, \citenamefont
  {{Arnold}}, \citenamefont {{Masson}}, \citenamefont {{Parkins}},\ and\
  \citenamefont {{Barrett}}}]{Zhiqiang2017}%
  \BibitemOpen
  \bibfield  {author} {\bibinfo {author} {\bibfnamefont {Z.}~\bibnamefont
  {{Zhiqiang}}}, \bibinfo {author} {\bibfnamefont {C.~H.}\ \bibnamefont
  {{Lee}}}, \bibinfo {author} {\bibfnamefont {R.}~\bibnamefont {{Kumar}}},
  \bibinfo {author} {\bibfnamefont {K.~J.}\ \bibnamefont {{Arnold}}}, \bibinfo
  {author} {\bibfnamefont {S.~J.}\ \bibnamefont {{Masson}}}, \bibinfo {author}
  {\bibfnamefont {A.~S.}\ \bibnamefont {{Parkins}}}, \ and\ \bibinfo {author}
  {\bibfnamefont {M.~D.}\ \bibnamefont {{Barrett}}},\ }\bibfield  {title}
  {\enquote {\bibinfo {title} {{Nonequilibrium phase transition in a spin-1
  Dicke model}},}\ }\href {\doibase 10.1364/OPTICA.4.000424} {\bibfield
  {journal} {\bibinfo  {journal} {Optica}\ }\textbf {\bibinfo {volume} {4}},\
  \bibinfo {pages} {424} (\bibinfo {year} {2017})}\BibitemShut {NoStop}%
\bibitem [{\citenamefont {Soriente}\ \emph {et~al.}(2018)\citenamefont
  {Soriente}, \citenamefont {Donner}, \citenamefont {Chitra},\ and\
  \citenamefont {Zilberberg}}]{Soriente2018}%
  \BibitemOpen
  \bibfield  {author} {\bibinfo {author} {\bibfnamefont {M.}~\bibnamefont
  {Soriente}}, \bibinfo {author} {\bibfnamefont {T.}~\bibnamefont {Donner}},
  \bibinfo {author} {\bibfnamefont {R.}~\bibnamefont {Chitra}}, \ and\ \bibinfo
  {author} {\bibfnamefont {O.}~\bibnamefont {Zilberberg}},\ }\bibfield  {title}
  {\enquote {\bibinfo {title} {{Dissipation-Induced Anomalous Multicritical
  Phenomena}},}\ }\href {\doibase 10.1103/PhysRevLett.120.183603} {\bibfield
  {journal} {\bibinfo  {journal} {Phys. Rev. Lett.}\ }\textbf {\bibinfo
  {volume} {120}},\ \bibinfo {pages} {183603} (\bibinfo {year}
  {2018})}\BibitemShut {NoStop}%
\bibitem [{\citenamefont {Chiacchio}\ and\ \citenamefont
  {Nunnenkamp}(2019)}]{Chiacchio2019}%
  \BibitemOpen
  \bibfield  {author} {\bibinfo {author} {\bibfnamefont {E.~I.~R.}\
  \bibnamefont {Chiacchio}}\ and\ \bibinfo {author} {\bibfnamefont
  {A.}~\bibnamefont {Nunnenkamp}},\ }\bibfield  {title} {\enquote {\bibinfo
  {title} {Dissipation-induced instabilities of a spinor bose-einstein
  condensate inside an optical cavity},}\ }\href {\doibase
  10.1103/PhysRevLett.122.193605} {\bibfield  {journal} {\bibinfo  {journal}
  {Phys. Rev. Lett.}\ }\textbf {\bibinfo {volume} {122}},\ \bibinfo {pages}
  {193605} (\bibinfo {year} {2019})}\BibitemShut {NoStop}%
\bibitem [{\citenamefont {Bu\ifmmode~\check{c}\else \v{c}\fi{}a}\ and\
  \citenamefont {Jaksch}(2019)}]{Buca2019}%
  \BibitemOpen
  \bibfield  {author} {\bibinfo {author} {\bibfnamefont {B.}~\bibnamefont
  {Bu\ifmmode~\check{c}\else \v{c}\fi{}a}}\ and\ \bibinfo {author}
  {\bibfnamefont {D.}~\bibnamefont {Jaksch}},\ }\bibfield  {title} {\enquote
  {\bibinfo {title} {Dissipation induced nonstationarity in a quantum gas},}\
  }\href {\doibase 10.1103/PhysRevLett.123.260401} {\bibfield  {journal}
  {\bibinfo  {journal} {Phys. Rev. Lett.}\ }\textbf {\bibinfo {volume} {123}},\
  \bibinfo {pages} {260401} (\bibinfo {year} {2019})}\BibitemShut {NoStop}%
\bibitem [{\citenamefont {Stitely}\ \emph {et~al.}(2020)\citenamefont
  {Stitely}, \citenamefont {Masson}, \citenamefont {Giraldo}, \citenamefont
  {Krauskopf},\ and\ \citenamefont {Parkins}}]{Stitely2020}%
  \BibitemOpen
  \bibfield  {author} {\bibinfo {author} {\bibfnamefont {K.~C.}\ \bibnamefont
  {Stitely}}, \bibinfo {author} {\bibfnamefont {S.~J.}\ \bibnamefont {Masson}},
  \bibinfo {author} {\bibfnamefont {A.}~\bibnamefont {Giraldo}}, \bibinfo
  {author} {\bibfnamefont {B.}~\bibnamefont {Krauskopf}}, \ and\ \bibinfo
  {author} {\bibfnamefont {S.}~\bibnamefont {Parkins}},\ }\bibfield  {title}
  {\enquote {\bibinfo {title} {{Superradiant switching, quantum hysteresis, and
  oscillations in a generalized Dicke model}},}\ }\href {\doibase
  10.1103/PhysRevA.102.063702} {\bibfield  {journal} {\bibinfo  {journal}
  {Phys. Rev. A}\ }\textbf {\bibinfo {volume} {102}},\ \bibinfo {pages}
  {063702} (\bibinfo {year} {2020})}\BibitemShut {NoStop}%
\bibitem [{\citenamefont {Habibian}\ \emph {et~al.}(2013)\citenamefont
  {Habibian}, \citenamefont {Winter}, \citenamefont {Paganelli}, \citenamefont
  {Rieger},\ and\ \citenamefont {Morigi}}]{Habibian2013}%
  \BibitemOpen
  \bibfield  {author} {\bibinfo {author} {\bibfnamefont {H.}~\bibnamefont
  {Habibian}}, \bibinfo {author} {\bibfnamefont {A.}~\bibnamefont {Winter}},
  \bibinfo {author} {\bibfnamefont {S.}~\bibnamefont {Paganelli}}, \bibinfo
  {author} {\bibfnamefont {H.}~\bibnamefont {Rieger}}, \ and\ \bibinfo {author}
  {\bibfnamefont {G.}~\bibnamefont {Morigi}},\ }\bibfield  {title} {\enquote
  {\bibinfo {title} {{Bose-Glass Phases of Ultracold Atoms due to Cavity
  Backaction}},}\ }\href {\doibase 10.1103/PhysRevLett.110.075304} {\bibfield
  {journal} {\bibinfo  {journal} {Phys. Rev. Lett.}\ }\textbf {\bibinfo
  {volume} {110}},\ \bibinfo {pages} {075304} (\bibinfo {year}
  {2013})}\BibitemShut {NoStop}%
\bibitem [{\citenamefont {Kollath}\ \emph {et~al.}(2016)\citenamefont
  {Kollath}, \citenamefont {Sheikhan}, \citenamefont {Wolff},\ and\
  \citenamefont {Brennecke}}]{Kollath2016}%
  \BibitemOpen
  \bibfield  {author} {\bibinfo {author} {\bibfnamefont {C.}~\bibnamefont
  {Kollath}}, \bibinfo {author} {\bibfnamefont {A.}~\bibnamefont {Sheikhan}},
  \bibinfo {author} {\bibfnamefont {S.}~\bibnamefont {Wolff}}, \ and\ \bibinfo
  {author} {\bibfnamefont {F.}~\bibnamefont {Brennecke}},\ }\bibfield  {title}
  {\enquote {\bibinfo {title} {{Ultracold Fermions in a Cavity-Induced
  Artificial Magnetic Field}},}\ }\href {\doibase
  10.1103/PhysRevLett.116.060401} {\bibfield  {journal} {\bibinfo  {journal}
  {Phys. Rev. Lett.}\ }\textbf {\bibinfo {volume} {116}},\ \bibinfo {pages}
  {060401} (\bibinfo {year} {2016})}\BibitemShut {NoStop}%
\bibitem [{\citenamefont {Mivehvar}\ \emph {et~al.}(2017)\citenamefont
  {Mivehvar}, \citenamefont {Piazza},\ and\ \citenamefont
  {Ritsch}}]{Mivehvar2017}%
  \BibitemOpen
  \bibfield  {author} {\bibinfo {author} {\bibfnamefont {F.}~\bibnamefont
  {Mivehvar}}, \bibinfo {author} {\bibfnamefont {F.}~\bibnamefont {Piazza}}, \
  and\ \bibinfo {author} {\bibfnamefont {H.}~\bibnamefont {Ritsch}},\
  }\bibfield  {title} {\enquote {\bibinfo {title} {{Disorder-Driven Density and
  Spin Self-Ordering of a Bose-Einstein Condensate in a Cavity}},}\ }\href
  {\doibase 10.1103/PhysRevLett.119.063602} {\bibfield  {journal} {\bibinfo
  {journal} {Phys. Rev. Lett.}\ }\textbf {\bibinfo {volume} {119}},\ \bibinfo
  {pages} {063602} (\bibinfo {year} {2017})}\BibitemShut {NoStop}%
\bibitem [{\citenamefont {{Vaidya}}\ \emph {et~al.}(2018)\citenamefont
  {{Vaidya}}, \citenamefont {{Guo}}, \citenamefont {{Kroeze}}, \citenamefont
  {{Ballantine}}, \citenamefont {{Koll{\'a}r}}, \citenamefont {{Keeling}},\
  and\ \citenamefont {{Lev}}}]{Vaidya2018}%
  \BibitemOpen
  \bibfield  {author} {\bibinfo {author} {\bibfnamefont {V.~D.}\ \bibnamefont
  {{Vaidya}}}, \bibinfo {author} {\bibfnamefont {Y.}~\bibnamefont {{Guo}}},
  \bibinfo {author} {\bibfnamefont {R.~M.}\ \bibnamefont {{Kroeze}}}, \bibinfo
  {author} {\bibfnamefont {K.~E.}\ \bibnamefont {{Ballantine}}}, \bibinfo
  {author} {\bibfnamefont {A.~J.}\ \bibnamefont {{Koll{\'a}r}}}, \bibinfo
  {author} {\bibfnamefont {J.}~\bibnamefont {{Keeling}}}, \ and\ \bibinfo
  {author} {\bibfnamefont {B.~L.}\ \bibnamefont {{Lev}}},\ }\bibfield  {title}
  {\enquote {\bibinfo {title} {{Tunable-Range, Photon-Mediated Atomic
  Interactions in Multimode Cavity QED}},}\ }\href {\doibase
  10.1103/PhysRevX.8.011002} {\bibfield  {journal} {\bibinfo  {journal}
  {Physical Review X}\ }\textbf {\bibinfo {volume} {8}},\ \bibinfo {eid}
  {011002} (\bibinfo {year} {2018})}\BibitemShut {NoStop}%
\bibitem [{\citenamefont {Landini}\ \emph {et~al.}(2018)\citenamefont
  {Landini}, \citenamefont {Dogra}, \citenamefont {Kroeger}, \citenamefont
  {Hruby}, \citenamefont {Donner},\ and\ \citenamefont
  {Esslinger}}]{Landini2018}%
  \BibitemOpen
  \bibfield  {author} {\bibinfo {author} {\bibfnamefont {M.}~\bibnamefont
  {Landini}}, \bibinfo {author} {\bibfnamefont {N.}~\bibnamefont {Dogra}},
  \bibinfo {author} {\bibfnamefont {K.}~\bibnamefont {Kroeger}}, \bibinfo
  {author} {\bibfnamefont {L.}~\bibnamefont {Hruby}}, \bibinfo {author}
  {\bibfnamefont {T.}~\bibnamefont {Donner}}, \ and\ \bibinfo {author}
  {\bibfnamefont {T.}~\bibnamefont {Esslinger}},\ }\bibfield  {title} {\enquote
  {\bibinfo {title} {Formation of a spin texture in a quantum gas coupled to a
  cavity},}\ }\href {\doibase 10.1103/PhysRevLett.120.223602} {\bibfield
  {journal} {\bibinfo  {journal} {Phys. Rev. Lett.}\ }\textbf {\bibinfo
  {volume} {120}},\ \bibinfo {pages} {223602} (\bibinfo {year}
  {2018})}\BibitemShut {NoStop}%
\bibitem [{\citenamefont {{Dogra}}\ \emph {et~al.}(2019)\citenamefont
  {{Dogra}}, \citenamefont {{Landini}}, \citenamefont {{Kroeger}},
  \citenamefont {{Hruby}}, \citenamefont {{Donner}},\ and\ \citenamefont
  {{Esslinger}}}]{Dogra2019}%
  \BibitemOpen
  \bibfield  {author} {\bibinfo {author} {\bibfnamefont {N.}~\bibnamefont
  {{Dogra}}}, \bibinfo {author} {\bibfnamefont {M.}~\bibnamefont {{Landini}}},
  \bibinfo {author} {\bibfnamefont {K.}~\bibnamefont {{Kroeger}}}, \bibinfo
  {author} {\bibfnamefont {L.}~\bibnamefont {{Hruby}}}, \bibinfo {author}
  {\bibfnamefont {T.}~\bibnamefont {{Donner}}}, \ and\ \bibinfo {author}
  {\bibfnamefont {T.}~\bibnamefont {{Esslinger}}},\ }\bibfield  {title}
  {\enquote {\bibinfo {title} {{Dissipation-induced structural instability and
  chiral dynamics in a quantum gas}},}\ }\href {\doibase
  10.1126/science.aaw4465} {\bibfield  {journal} {\bibinfo  {journal}
  {Science}\ }\textbf {\bibinfo {volume} {366}},\ \bibinfo {pages} {1496--1499}
  (\bibinfo {year} {2019})}\BibitemShut {NoStop}%
\bibitem [{\citenamefont {Bentsen}\ \emph {et~al.}(2019)\citenamefont
  {Bentsen}, \citenamefont {Potirniche}, \citenamefont {Bulchandani},
  \citenamefont {Scaffidi}, \citenamefont {Cao}, \citenamefont {Qi},
  \citenamefont {Schleier-Smith},\ and\ \citenamefont {Altman}}]{Bentsen2019}%
  \BibitemOpen
  \bibfield  {author} {\bibinfo {author} {\bibfnamefont {G.}~\bibnamefont
  {Bentsen}}, \bibinfo {author} {\bibfnamefont {I.-D.}\ \bibnamefont
  {Potirniche}}, \bibinfo {author} {\bibfnamefont {V.~B.}\ \bibnamefont
  {Bulchandani}}, \bibinfo {author} {\bibfnamefont {T.}~\bibnamefont
  {Scaffidi}}, \bibinfo {author} {\bibfnamefont {X.}~\bibnamefont {Cao}},
  \bibinfo {author} {\bibfnamefont {X.-L.}\ \bibnamefont {Qi}}, \bibinfo
  {author} {\bibfnamefont {M.}~\bibnamefont {Schleier-Smith}}, \ and\ \bibinfo
  {author} {\bibfnamefont {E.}~\bibnamefont {Altman}},\ }\bibfield  {title}
  {\enquote {\bibinfo {title} {{Integrable and Chaotic Dynamics of Spins
  Coupled to an Optical Cavity}},}\ }\href {\doibase 10.1103/PhysRevX.9.041011}
  {\bibfield  {journal} {\bibinfo  {journal} {Phys. Rev. X}\ }\textbf {\bibinfo
  {volume} {9}},\ \bibinfo {pages} {041011} (\bibinfo {year}
  {2019})}\BibitemShut {NoStop}%
\bibitem [{\citenamefont {J\"ager}\ \emph {et~al.}(2020)\citenamefont
  {J\"ager}, \citenamefont {Holland},\ and\ \citenamefont
  {Morigi}}]{Jager2020}%
  \BibitemOpen
  \bibfield  {author} {\bibinfo {author} {\bibfnamefont {S.~B.}\ \bibnamefont
  {J\"ager}}, \bibinfo {author} {\bibfnamefont {M.~J.}\ \bibnamefont
  {Holland}}, \ and\ \bibinfo {author} {\bibfnamefont {G.}~\bibnamefont
  {Morigi}},\ }\bibfield  {title} {\enquote {\bibinfo {title} {Superradiant
  optomechanical phases of cold atomic gases in optical resonators},}\ }\href
  {\doibase 10.1103/PhysRevA.101.023616} {\bibfield  {journal} {\bibinfo
  {journal} {Phys. Rev. A}\ }\textbf {\bibinfo {volume} {101}},\ \bibinfo
  {pages} {023616} (\bibinfo {year} {2020})}\BibitemShut {NoStop}%
\bibitem [{\citenamefont {Boller}\ \emph {et~al.}(1991)\citenamefont {Boller},
  \citenamefont {Imamo\ifmmode~\breve{g}\else \u{g}\fi{}lu},\ and\
  \citenamefont {Harris}}]{PhysRevLettBoller}%
  \BibitemOpen
  \bibfield  {author} {\bibinfo {author} {\bibfnamefont {K.-J.}\ \bibnamefont
  {Boller}}, \bibinfo {author} {\bibfnamefont {A.}~\bibnamefont
  {Imamo\ifmmode~\breve{g}\else \u{g}\fi{}lu}}, \ and\ \bibinfo {author}
  {\bibfnamefont {S.~E.}\ \bibnamefont {Harris}},\ }\bibfield  {title}
  {\enquote {\bibinfo {title} {Observation of electromagnetically induced
  transparency},}\ }\href {\doibase 10.1103/PhysRevLett.66.2593} {\bibfield
  {journal} {\bibinfo  {journal} {Phys. Rev. Lett.}\ }\textbf {\bibinfo
  {volume} {66}},\ \bibinfo {pages} {2593--2596} (\bibinfo {year}
  {1991})}\BibitemShut {NoStop}%
\bibitem [{\citenamefont {Fleischhauer}\ \emph {et~al.}(2005)\citenamefont
  {Fleischhauer}, \citenamefont {Imamoglu},\ and\ \citenamefont
  {Marangos}}]{RevModPhysFleischhauer}%
  \BibitemOpen
  \bibfield  {author} {\bibinfo {author} {\bibfnamefont {M.}~\bibnamefont
  {Fleischhauer}}, \bibinfo {author} {\bibfnamefont {A.}~\bibnamefont
  {Imamoglu}}, \ and\ \bibinfo {author} {\bibfnamefont {J.~P.}\ \bibnamefont
  {Marangos}},\ }\bibfield  {title} {\enquote {\bibinfo {title}
  {Electromagnetically induced transparency: Optics in coherent media},}\
  }\href {\doibase 10.1103/RevModPhys.77.633} {\bibfield  {journal} {\bibinfo
  {journal} {Rev. Mod. Phys.}\ }\textbf {\bibinfo {volume} {77}},\ \bibinfo
  {pages} {633--673} (\bibinfo {year} {2005})}\BibitemShut {NoStop}%
\bibitem [{\citenamefont {Scully}\ \emph {et~al.}(1989)\citenamefont {Scully},
  \citenamefont {Zhu},\ and\ \citenamefont {Gavrielides}}]{PhysRevLettScully}%
  \BibitemOpen
  \bibfield  {author} {\bibinfo {author} {\bibfnamefont {M.~O.}\ \bibnamefont
  {Scully}}, \bibinfo {author} {\bibfnamefont {S.-Y.}\ \bibnamefont {Zhu}}, \
  and\ \bibinfo {author} {\bibfnamefont {A.}~\bibnamefont {Gavrielides}},\
  }\bibfield  {title} {\enquote {\bibinfo {title} {Degenerate quantum-beat
  laser: Lasing without inversion and inversion without lasing},}\ }\href
  {\doibase 10.1103/PhysRevLett.62.2813} {\bibfield  {journal} {\bibinfo
  {journal} {Phys. Rev. Lett.}\ }\textbf {\bibinfo {volume} {62}},\ \bibinfo
  {pages} {2813--2816} (\bibinfo {year} {1989})}\BibitemShut {NoStop}%
\bibitem [{\citenamefont {Mompart}\ and\ \citenamefont
  {Corbal{\'{a}}n}(2000)}]{Mompart_2000}%
  \BibitemOpen
  \bibfield  {author} {\bibinfo {author} {\bibfnamefont {J}~\bibnamefont
  {Mompart}}\ and\ \bibinfo {author} {\bibfnamefont {R}~\bibnamefont
  {Corbal{\'{a}}n}},\ }\bibfield  {title} {\enquote {\bibinfo {title} {Lasing
  without inversion},}\ }\href {\doibase 10.1088/1464-4266/2/3/201} {\bibfield
  {journal} {\bibinfo  {journal} {J. Opt. B: Quantum Semiclass. Opt.}\ }\textbf
  {\bibinfo {volume} {2}},\ \bibinfo {pages} {R7--R24} (\bibinfo {year}
  {2000})}\BibitemShut {NoStop}%
\bibitem [{\citenamefont {Gaubatz}\ \emph {et~al.}(1990)\citenamefont
  {Gaubatz}, \citenamefont {Rudecki}, \citenamefont {Schiemann},\ and\
  \citenamefont {Bergmann}}]{Gaubatz}%
  \BibitemOpen
  \bibfield  {author} {\bibinfo {author} {\bibfnamefont {U.}~\bibnamefont
  {Gaubatz}}, \bibinfo {author} {\bibfnamefont {P.}~\bibnamefont {Rudecki}},
  \bibinfo {author} {\bibfnamefont {S.}~\bibnamefont {Schiemann}}, \ and\
  \bibinfo {author} {\bibfnamefont {K.}~\bibnamefont {Bergmann}},\ }\bibfield
  {title} {\enquote {\bibinfo {title} {Population transfer between molecular
  vibrational levels by stimulated raman scattering with partially overlapping
  laser fields. a new concept and experimental results},}\ }\href {\doibase
  10.1063/1.458514} {\bibfield  {journal} {\bibinfo  {journal} {J. Chem. Phys}\
  }\textbf {\bibinfo {volume} {92}},\ \bibinfo {pages} {5363--5376} (\bibinfo
  {year} {1990})}\BibitemShut {NoStop}%
\bibitem [{\citenamefont {Vitanov}\ \emph {et~al.}(2017)\citenamefont
  {Vitanov}, \citenamefont {Rangelov}, \citenamefont {Shore},\ and\
  \citenamefont {Bergmann}}]{RevModPhysVitanov}%
  \BibitemOpen
  \bibfield  {author} {\bibinfo {author} {\bibfnamefont {N.~V.}\ \bibnamefont
  {Vitanov}}, \bibinfo {author} {\bibfnamefont {A.~A.}\ \bibnamefont
  {Rangelov}}, \bibinfo {author} {\bibfnamefont {B.~W.}\ \bibnamefont {Shore}},
  \ and\ \bibinfo {author} {\bibfnamefont {K.}~\bibnamefont {Bergmann}},\
  }\bibfield  {title} {\enquote {\bibinfo {title} {Stimulated raman adiabatic
  passage in physics, chemistry, and beyond},}\ }\href {\doibase
  10.1103/RevModPhys.89.015006} {\bibfield  {journal} {\bibinfo  {journal}
  {Rev. Mod. Phys.}\ }\textbf {\bibinfo {volume} {89}},\ \bibinfo {pages}
  {015006} (\bibinfo {year} {2017})}\BibitemShut {NoStop}%
\bibitem [{\citenamefont {{Sung}}\ and\ \citenamefont
  {{Bowden}}(1979)}]{Sung1979}%
  \BibitemOpen
  \bibfield  {author} {\bibinfo {author} {\bibfnamefont {C.~C.}\ \bibnamefont
  {{Sung}}}\ and\ \bibinfo {author} {\bibfnamefont {C.~M.}\ \bibnamefont
  {{Bowden}}},\ }\bibfield  {title} {\enquote {\bibinfo {title} {{Phase
  transition in the multimode two- and three-level Dicke model (Green's
  function method)}},}\ }\href {\doibase 10.1088/0305-4470/12/11/035}
  {\bibfield  {journal} {\bibinfo  {journal} {J. Phys. A: Math. Gen.}\ }\textbf
  {\bibinfo {volume} {12}},\ \bibinfo {pages} {2273} (\bibinfo {year}
  {1979})}\BibitemShut {NoStop}%
\bibitem [{\citenamefont {{Crubellier}}\ \emph {et~al.}(1985)\citenamefont
  {{Crubellier}}, \citenamefont {{Liberman}}, \citenamefont {{Pavolini}},\ and\
  \citenamefont {{Pillet}}}]{Crubellier1985}%
  \BibitemOpen
  \bibfield  {author} {\bibinfo {author} {\bibfnamefont {A.}~\bibnamefont
  {{Crubellier}}}, \bibinfo {author} {\bibfnamefont {S.}~\bibnamefont
  {{Liberman}}}, \bibinfo {author} {\bibfnamefont {D.}~\bibnamefont
  {{Pavolini}}}, \ and\ \bibinfo {author} {\bibfnamefont {P.}~\bibnamefont
  {{Pillet}}},\ }\bibfield  {title} {\enquote {\bibinfo {title} {{Superradiance
  and subradiance. I. Interatomic interference and symmetry properties in
  three-level systems}},}\ }\href {\doibase 10.1088/0022-3700/18/18/022}
  {\bibfield  {journal} {\bibinfo  {journal} {J. Phys. B: At. Mol. Phys.}\
  }\textbf {\bibinfo {volume} {18}},\ \bibinfo {pages} {3811--3833} (\bibinfo
  {year} {1985})}\BibitemShut {NoStop}%
\bibitem [{\citenamefont {{Crubellier}}\ and\ \citenamefont
  {{Pavolini}}(1986)}]{Crubellier1986}%
  \BibitemOpen
  \bibfield  {author} {\bibinfo {author} {\bibfnamefont {A.}~\bibnamefont
  {{Crubellier}}}\ and\ \bibinfo {author} {\bibfnamefont {D.}~\bibnamefont
  {{Pavolini}}},\ }\bibfield  {title} {\enquote {\bibinfo {title}
  {{Superradiance and subradiance. II. Atomic systems with degenerate
  transitions}},}\ }\href {\doibase 10.1088/0022-3700/19/14/008} {\bibfield
  {journal} {\bibinfo  {journal} {J. Phys. B: At. Mol. Phys.}\ }\textbf
  {\bibinfo {volume} {19}},\ \bibinfo {pages} {2109--2138} (\bibinfo {year}
  {1986})}\BibitemShut {NoStop}%
\bibitem [{\citenamefont {Cola}\ \emph {et~al.}(2009)\citenamefont {Cola},
  \citenamefont {Bigerni},\ and\ \citenamefont {Piovella}}]{Cola2009}%
  \BibitemOpen
  \bibfield  {author} {\bibinfo {author} {\bibfnamefont {M.~M.}\ \bibnamefont
  {Cola}}, \bibinfo {author} {\bibfnamefont {D.}~\bibnamefont {Bigerni}}, \
  and\ \bibinfo {author} {\bibfnamefont {N.}~\bibnamefont {Piovella}},\
  }\bibfield  {title} {\enquote {\bibinfo {title} {Recoil-induced subradiance
  in an ultracold atomic gas},}\ }\href {\doibase 10.1103/PhysRevA.79.053622}
  {\bibfield  {journal} {\bibinfo  {journal} {Phys. Rev. A}\ }\textbf {\bibinfo
  {volume} {79}},\ \bibinfo {pages} {053622} (\bibinfo {year}
  {2009})}\BibitemShut {NoStop}%
\bibitem [{\citenamefont {Wolf}\ \emph {et~al.}(2018)\citenamefont {Wolf},
  \citenamefont {Schuster}, \citenamefont {Schmidt}, \citenamefont {Slama},\
  and\ \citenamefont {Zimmermann}}]{Wolf2018}%
  \BibitemOpen
  \bibfield  {author} {\bibinfo {author} {\bibfnamefont {P.}~\bibnamefont
  {Wolf}}, \bibinfo {author} {\bibfnamefont {S.~C.}\ \bibnamefont {Schuster}},
  \bibinfo {author} {\bibfnamefont {D.}~\bibnamefont {Schmidt}}, \bibinfo
  {author} {\bibfnamefont {S.}~\bibnamefont {Slama}}, \ and\ \bibinfo {author}
  {\bibfnamefont {C.}~\bibnamefont {Zimmermann}},\ }\bibfield  {title}
  {\enquote {\bibinfo {title} {{Observation of Subradiant Atomic Momentum
  States with Bose-Einstein Condensates in a Recoil Resolving Optical Ring
  Resonator}},}\ }\href {\doibase 10.1103/PhysRevLett.121.173602} {\bibfield
  {journal} {\bibinfo  {journal} {Phys. Rev. Lett.}\ }\textbf {\bibinfo
  {volume} {121}},\ \bibinfo {pages} {173602} (\bibinfo {year}
  {2018})}\BibitemShut {NoStop}%
\bibitem [{\citenamefont {Georgi}(2018)}]{georgi2018lie}%
  \BibitemOpen
  \bibfield  {author} {\bibinfo {author} {\bibfnamefont {H.}~\bibnamefont
  {Georgi}},\ }\href {https://books.google.de/books?id=CUpaDwAAQBAJ} {\emph
  {\bibinfo {title} {{Lie Algebras In Particle Physics: from Isospin To Unified
  Theories}}}}\ (\bibinfo  {publisher} {CRC Press},\ \bibinfo {year}
  {2018})\BibitemShut {NoStop}%
\bibitem [{\citenamefont {Marciano}\ and\ \citenamefont
  {Pagels}(1978)}]{MARCIANO}%
  \BibitemOpen
  \bibfield  {author} {\bibinfo {author} {\bibfnamefont {W.}~\bibnamefont
  {Marciano}}\ and\ \bibinfo {author} {\bibfnamefont {H.}~\bibnamefont
  {Pagels}},\ }\bibfield  {title} {\enquote {\bibinfo {title} {Quantum
  chromodynamics},}\ }\href {\doibase
  https://doi.org/10.1016/0370-1573(78)90208-9} {\bibfield  {journal} {\bibinfo
   {journal} {Physics Reports}\ }\textbf {\bibinfo {volume} {36}},\ \bibinfo
  {pages} {137--276} (\bibinfo {year} {1978})}\BibitemShut {NoStop}%
\bibitem [{\citenamefont {Griffiths}(2008)}]{Griffiths2008}%
  \BibitemOpen
  \bibfield  {author} {\bibinfo {author} {\bibfnamefont {D.}~\bibnamefont
  {Griffiths}},\ }\href@noop {} {\emph {\bibinfo {title} {{Introduction to
  elementary particles}}}}\ (\bibinfo {year} {2008})\BibitemShut {NoStop}%
\bibitem [{\citenamefont {Cosme}\ \emph {et~al.}(2019)\citenamefont {Cosme},
  \citenamefont {Skulte},\ and\ \citenamefont {Mathey}}]{Cosme2019}%
  \BibitemOpen
  \bibfield  {author} {\bibinfo {author} {\bibfnamefont {J.~G.}\ \bibnamefont
  {Cosme}}, \bibinfo {author} {\bibfnamefont {J.}~\bibnamefont {Skulte}}, \
  and\ \bibinfo {author} {\bibfnamefont {L.}~\bibnamefont {Mathey}},\
  }\bibfield  {title} {\enquote {\bibinfo {title} {{Time crystals in a shaken
  atom-cavity system}},}\ }\href {\doibase 10.1103/PhysRevA.100.053615}
  {\bibfield  {journal} {\bibinfo  {journal} {Phys. Rev. A}\ }\textbf {\bibinfo
  {volume} {100}},\ \bibinfo {pages} {053615} (\bibinfo {year}
  {2019})}\BibitemShut {NoStop}%
\bibitem [{joi()}]{joint}%
  \BibitemOpen
  \href@noop {} {}\bibinfo {note} {{See P. {Kongkhambut} et al. for
  details}}\BibitemShut {NoStop}%
\bibitem [{\citenamefont {Sakurai}\ and\ \citenamefont
  {Napolitano}(2017)}]{sakurai_napolitano_2017}%
  \BibitemOpen
  \bibfield  {author} {\bibinfo {author} {\bibfnamefont {J.~J.}\ \bibnamefont
  {Sakurai}}\ and\ \bibinfo {author} {\bibfnamefont {Jim}\ \bibnamefont
  {Napolitano}},\ }\href {\doibase 10.1017/9781108499996} {\emph {\bibinfo
  {title} {Modern Quantum Mechanics}}},\ \bibinfo {edition} {2nd}\ ed.\
  (\bibinfo  {publisher} {Cambridge University Press},\ \bibinfo {year}
  {2017})\BibitemShut {NoStop}%
\bibitem [{\citenamefont {Baksic}\ \emph {et~al.}(2013)\citenamefont {Baksic},
  \citenamefont {Nataf},\ and\ \citenamefont {Ciuti}}]{Baksic2013}%
  \BibitemOpen
  \bibfield  {author} {\bibinfo {author} {\bibfnamefont {A.}~\bibnamefont
  {Baksic}}, \bibinfo {author} {\bibfnamefont {P.}~\bibnamefont {Nataf}}, \
  and\ \bibinfo {author} {\bibfnamefont {C.}~\bibnamefont {Ciuti}},\ }\bibfield
   {title} {\enquote {\bibinfo {title} {Superradiant phase transitions with
  three-level systems},}\ }\href {\doibase 10.1103/PhysRevA.87.023813}
  {\bibfield  {journal} {\bibinfo  {journal} {Phys. Rev. A}\ }\textbf {\bibinfo
  {volume} {87}},\ \bibinfo {pages} {023813} (\bibinfo {year}
  {2013})}\BibitemShut {NoStop}%
\bibitem [{\citenamefont {Dimer}\ \emph {et~al.}(2007)\citenamefont {Dimer},
  \citenamefont {Estienne}, \citenamefont {Parkins},\ and\ \citenamefont
  {Carmichael}}]{Dimer2007}%
  \BibitemOpen
  \bibfield  {author} {\bibinfo {author} {\bibfnamefont {F.}~\bibnamefont
  {Dimer}}, \bibinfo {author} {\bibfnamefont {B.}~\bibnamefont {Estienne}},
  \bibinfo {author} {\bibfnamefont {A.~S.}\ \bibnamefont {Parkins}}, \ and\
  \bibinfo {author} {\bibfnamefont {H.~J.}\ \bibnamefont {Carmichael}},\
  }\bibfield  {title} {\enquote {\bibinfo {title} {Proposed realization of the
  dicke-model quantum phase transition in an optical cavity qed system},}\
  }\href {\doibase 10.1103/PhysRevA.75.013804} {\bibfield  {journal} {\bibinfo
  {journal} {Phys. Rev. A}\ }\textbf {\bibinfo {volume} {75}},\ \bibinfo
  {pages} {013804} (\bibinfo {year} {2007})}\BibitemShut {NoStop}%
\bibitem [{\citenamefont {Kirton}\ \emph {et~al.}(2019)\citenamefont {Kirton},
  \citenamefont {Roses}, \citenamefont {Keeling},\ and\ \citenamefont
  {Dalla~Torre}}]{Kirton2019}%
  \BibitemOpen
  \bibfield  {author} {\bibinfo {author} {\bibfnamefont {P.}~\bibnamefont
  {Kirton}}, \bibinfo {author} {\bibfnamefont {M.~M.}\ \bibnamefont {Roses}},
  \bibinfo {author} {\bibfnamefont {J.}~\bibnamefont {Keeling}}, \ and\
  \bibinfo {author} {\bibfnamefont {E.~G.}\ \bibnamefont {Dalla~Torre}},\
  }\bibfield  {title} {\enquote {\bibinfo {title} {{Introduction to the Dicke
  Model: From Equilibrium to Nonequilibrium, and Vice Versa}},}\ }\href
  {\doibase 10.1002/qute.201800043} {\bibfield  {journal} {\bibinfo  {journal}
  {Adv. Quantum Technol.}\ }\textbf {\bibinfo {volume} {2}},\ \bibinfo {pages}
  {1800043} (\bibinfo {year} {2019})}\BibitemShut {NoStop}%
\bibitem [{\citenamefont {{Mivehvar}}\ \emph {et~al.}(2021)\citenamefont
  {{Mivehvar}}, \citenamefont {{Piazza}}, \citenamefont {{Donner}},\ and\
  \citenamefont {{Ritsch}}}]{Mivehvar2021}%
  \BibitemOpen
  \bibfield  {author} {\bibinfo {author} {\bibfnamefont {F.}~\bibnamefont
  {{Mivehvar}}}, \bibinfo {author} {\bibfnamefont {F.}~\bibnamefont
  {{Piazza}}}, \bibinfo {author} {\bibfnamefont {T.}~\bibnamefont {{Donner}}},
  \ and\ \bibinfo {author} {\bibfnamefont {H.}~\bibnamefont {{Ritsch}}},\
  }\bibfield  {title} {\enquote {\bibinfo {title} {{Cavity QED with Quantum
  Gases: New Paradigms in Many-Body Physics}},}\ }\href@noop {} {\bibfield
  {journal} {\bibinfo  {journal} {arXiv e-prints}\ ,\ \bibinfo {eid}
  {arXiv:2102.04473}} (\bibinfo {year} {2021})},\ \Eprint
  {http://arxiv.org/abs/2102.04473} {arXiv:2102.04473} \BibitemShut {NoStop}%
\bibitem [{\citenamefont {Ritsch}\ \emph {et~al.}(2013)\citenamefont {Ritsch},
  \citenamefont {Domokos}, \citenamefont {Brennecke},\ and\ \citenamefont
  {Esslinger}}]{Ritsch2013}%
  \BibitemOpen
  \bibfield  {author} {\bibinfo {author} {\bibfnamefont {H.}~\bibnamefont
  {Ritsch}}, \bibinfo {author} {\bibfnamefont {P.}~\bibnamefont {Domokos}},
  \bibinfo {author} {\bibfnamefont {F.}~\bibnamefont {Brennecke}}, \ and\
  \bibinfo {author} {\bibfnamefont {T.}~\bibnamefont {Esslinger}},\ }\bibfield
  {title} {\enquote {\bibinfo {title} {Cold atoms in cavity-generated dynamical
  optical potentials},}\ }\href {\doibase 10.1103/RevModPhys.85.553} {\bibfield
   {journal} {\bibinfo  {journal} {Rev. Mod. Phys.}\ }\textbf {\bibinfo
  {volume} {85}},\ \bibinfo {pages} {553--601} (\bibinfo {year}
  {2013})}\BibitemShut {NoStop}%
\bibitem [{\citenamefont {{Cosme}}\ \emph {et~al.}(2018)\citenamefont
  {{Cosme}}, \citenamefont {{Georges}}, \citenamefont {{Hemmerich}},\ and\
  \citenamefont {{Mathey}}}]{Cosme2018}%
  \BibitemOpen
  \bibfield  {author} {\bibinfo {author} {\bibfnamefont {J.~G.}\ \bibnamefont
  {{Cosme}}}, \bibinfo {author} {\bibfnamefont {C.}~\bibnamefont {{Georges}}},
  \bibinfo {author} {\bibfnamefont {A.}~\bibnamefont {{Hemmerich}}}, \ and\
  \bibinfo {author} {\bibfnamefont {L.}~\bibnamefont {{Mathey}}},\ }\bibfield
  {title} {\enquote {\bibinfo {title} {{Dynamical Control of Order in a
  Cavity-BEC System}},}\ }\href {\doibase 10.1103/PhysRevLett.121.153001}
  {\bibfield  {journal} {\bibinfo  {journal} {Phys. Rev. Lett.}\ }\textbf
  {\bibinfo {volume} {121}},\ \bibinfo {pages} {153001} (\bibinfo {year}
  {2018})}\BibitemShut {NoStop}%
\bibitem [{\citenamefont {{Georges}}\ \emph {et~al.}(2018)\citenamefont
  {{Georges}}, \citenamefont {{Cosme}}, \citenamefont {{Mathey}},\ and\
  \citenamefont {{Hemmerich}}}]{Georges2018}%
  \BibitemOpen
  \bibfield  {author} {\bibinfo {author} {\bibfnamefont {C.}~\bibnamefont
  {{Georges}}}, \bibinfo {author} {\bibfnamefont {J.~G.}\ \bibnamefont
  {{Cosme}}}, \bibinfo {author} {\bibfnamefont {L.}~\bibnamefont {{Mathey}}}, \
  and\ \bibinfo {author} {\bibfnamefont {A.}~\bibnamefont {{Hemmerich}}},\
  }\bibfield  {title} {\enquote {\bibinfo {title} {{Light-Induced Coherence in
  an Atom-Cavity System}},}\ }\href {\doibase 10.1103/PhysRevLett.121.220405}
  {\bibfield  {journal} {\bibinfo  {journal} {Phys. Rev. Lett.}\ }\textbf
  {\bibinfo {volume} {121}},\ \bibinfo {pages} {220405} (\bibinfo {year}
  {2018})}\BibitemShut {NoStop}%
\bibitem [{\citenamefont {{Ke{\ss}ler}}\ \emph {et~al.}(2019)\citenamefont
  {{Ke{\ss}ler}}, \citenamefont {{Cosme}}, \citenamefont {{Hemmerling}},
  \citenamefont {{Mathey}},\ and\ \citenamefont {{Hemmerich}}}]{Kessler2019}%
  \BibitemOpen
  \bibfield  {author} {\bibinfo {author} {\bibfnamefont {H.}~\bibnamefont
  {{Ke{\ss}ler}}}, \bibinfo {author} {\bibfnamefont {J.~G.}\ \bibnamefont
  {{Cosme}}}, \bibinfo {author} {\bibfnamefont {M.}~\bibnamefont
  {{Hemmerling}}}, \bibinfo {author} {\bibfnamefont {L.}~\bibnamefont
  {{Mathey}}}, \ and\ \bibinfo {author} {\bibfnamefont {A.}~\bibnamefont
  {{Hemmerich}}},\ }\bibfield  {title} {\enquote {\bibinfo {title} {{Emergent
  limit cycles and time crystal dynamics in an atom-cavity system}},}\ }\href
  {\doibase 10.1103/PhysRevA.99.053605} {\bibfield  {journal} {\bibinfo
  {journal} {Phys. Rev. A}\ }\textbf {\bibinfo {volume} {99}},\ \bibinfo
  {pages} {053605} (\bibinfo {year} {2019})}\BibitemShut {NoStop}%
\bibitem [{\citenamefont {{Ke{\ss}ler}}\ \emph {et~al.}(2020)\citenamefont
  {{Ke{\ss}ler}}, \citenamefont {{Cosme}}, \citenamefont {{Georges}},
  \citenamefont {{Mathey}},\ and\ \citenamefont {{Hemmerich}}}]{Kessler2020}%
  \BibitemOpen
  \bibfield  {author} {\bibinfo {author} {\bibfnamefont {H.}~\bibnamefont
  {{Ke{\ss}ler}}}, \bibinfo {author} {\bibfnamefont {J.~G.}\ \bibnamefont
  {{Cosme}}}, \bibinfo {author} {\bibfnamefont {C.}~\bibnamefont {{Georges}}},
  \bibinfo {author} {\bibfnamefont {L.}~\bibnamefont {{Mathey}}}, \ and\
  \bibinfo {author} {\bibfnamefont {A.}~\bibnamefont {{Hemmerich}}},\
  }\bibfield  {title} {\enquote {\bibinfo {title} {{From a continuous to a
  discrete time crystal in a dissipative atom-cavity system}},}\ }\href
  {\doibase 10.1088/1367-2630/ab9fc0} {\bibfield  {journal} {\bibinfo
  {journal} {New J. Phys.}\ }\textbf {\bibinfo {volume} {22}},\ \bibinfo {eid}
  {085002} (\bibinfo {year} {2020})}\BibitemShut {NoStop}%
\bibitem [{\citenamefont {{Ke{\ss}ler}}\ \emph {et~al.}(2021)\citenamefont
  {{Ke{\ss}ler}}, \citenamefont {{Kongkhambut}}, \citenamefont {{Georges}},
  \citenamefont {{Mathey}}, \citenamefont {{Cosme}},\ and\ \citenamefont
  {{Hemmerich}}}]{Kessler2021}%
  \BibitemOpen
  \bibfield  {author} {\bibinfo {author} {\bibfnamefont {H.}~\bibnamefont
  {{Ke{\ss}ler}}}, \bibinfo {author} {\bibfnamefont {P.}~\bibnamefont
  {{Kongkhambut}}}, \bibinfo {author} {\bibfnamefont {C.}~\bibnamefont
  {{Georges}}}, \bibinfo {author} {\bibfnamefont {L.}~\bibnamefont {{Mathey}}},
  \bibinfo {author} {\bibfnamefont {J.~G.}\ \bibnamefont {{Cosme}}}, \ and\
  \bibinfo {author} {\bibfnamefont {A.}~\bibnamefont {{Hemmerich}}},\
  }\bibfield  {title} {\enquote {\bibinfo {title} {{Observation of a
  Dissipative Time Crystal}},}\ }\href {\doibase
  10.1103/PhysRevLett.127.043602} {\bibfield  {journal} {\bibinfo  {journal}
  {Phys. Rev. Lett.}\ }\textbf {\bibinfo {volume} {127}},\ \bibinfo {pages}
  {043602} (\bibinfo {year} {2021})}\BibitemShut {NoStop}%
\bibitem [{\citenamefont {{Georges}}\ \emph {et~al.}(2021)\citenamefont
  {{Georges}}, \citenamefont {{Cosme}}, \citenamefont {{Ke{\ss}ler}},
  \citenamefont {{Mathey}},\ and\ \citenamefont {{Hemmerich}}}]{Georges2021}%
  \BibitemOpen
  \bibfield  {author} {\bibinfo {author} {\bibfnamefont {C.}~\bibnamefont
  {{Georges}}}, \bibinfo {author} {\bibfnamefont {J.~G.}\ \bibnamefont
  {{Cosme}}}, \bibinfo {author} {\bibfnamefont {H.}~\bibnamefont
  {{Ke{\ss}ler}}}, \bibinfo {author} {\bibfnamefont {L.}~\bibnamefont
  {{Mathey}}}, \ and\ \bibinfo {author} {\bibfnamefont {A.}~\bibnamefont
  {{Hemmerich}}},\ }\bibfield  {title} {\enquote {\bibinfo {title} {{Dynamical
  density wave order in an atom-cavity system}},}\ }\href {\doibase
  10.1088/1367-2630/abdf9c} {\bibfield  {journal} {\bibinfo  {journal} {New J.
  Phys.}\ }\textbf {\bibinfo {volume} {23}},\ \bibinfo {eid} {023003} (\bibinfo
  {year} {2021})}\BibitemShut {NoStop}%
\bibitem [{\citenamefont {{Fausti}}\ \emph {et~al.}(2011)\citenamefont
  {{Fausti}}, \citenamefont {{Tobey}}, \citenamefont {{Dean}}, \citenamefont
  {{Kaiser}}, \citenamefont {{Dienst}}, \citenamefont {{Hoffmann}},
  \citenamefont {{Pyon}}, \citenamefont {{Takayama}}, \citenamefont
  {{Takagi}},\ and\ \citenamefont {{Cavalleri}}}]{Fausti2011}%
  \BibitemOpen
  \bibfield  {author} {\bibinfo {author} {\bibfnamefont {D.}~\bibnamefont
  {{Fausti}}}, \bibinfo {author} {\bibfnamefont {R.~I.}\ \bibnamefont
  {{Tobey}}}, \bibinfo {author} {\bibfnamefont {N.}~\bibnamefont {{Dean}}},
  \bibinfo {author} {\bibfnamefont {S.}~\bibnamefont {{Kaiser}}}, \bibinfo
  {author} {\bibfnamefont {A.}~\bibnamefont {{Dienst}}}, \bibinfo {author}
  {\bibfnamefont {M.~C.}\ \bibnamefont {{Hoffmann}}}, \bibinfo {author}
  {\bibfnamefont {S.}~\bibnamefont {{Pyon}}}, \bibinfo {author} {\bibfnamefont
  {T.}~\bibnamefont {{Takayama}}}, \bibinfo {author} {\bibfnamefont
  {H.}~\bibnamefont {{Takagi}}}, \ and\ \bibinfo {author} {\bibfnamefont
  {A.}~\bibnamefont {{Cavalleri}}},\ }\bibfield  {title} {\enquote {\bibinfo
  {title} {{Light-Induced Superconductivity in a Stripe-Ordered Cuprate}},}\
  }\href {\doibase 10.1126/science.1197294} {\bibfield  {journal} {\bibinfo
  {journal} {Science}\ }\textbf {\bibinfo {volume} {331}},\ \bibinfo {pages}
  {189} (\bibinfo {year} {2011})}\BibitemShut {NoStop}%
\bibitem [{\citenamefont {{Hu}}\ \emph {et~al.}(2014)\citenamefont {{Hu}},
  \citenamefont {{Kaiser}}, \citenamefont {{Nicoletti}}, \citenamefont
  {{Hunt}}, \citenamefont {{Gierz}}, \citenamefont {{Hoffmann}}, \citenamefont
  {{Le Tacon}}, \citenamefont {{Loew}}, \citenamefont {{Keimer}},\ and\
  \citenamefont {{Cavalleri}}}]{Hu2014}%
  \BibitemOpen
  \bibfield  {author} {\bibinfo {author} {\bibfnamefont {W.}~\bibnamefont
  {{Hu}}}, \bibinfo {author} {\bibfnamefont {S.}~\bibnamefont {{Kaiser}}},
  \bibinfo {author} {\bibfnamefont {D.}~\bibnamefont {{Nicoletti}}}, \bibinfo
  {author} {\bibfnamefont {C.~R.}\ \bibnamefont {{Hunt}}}, \bibinfo {author}
  {\bibfnamefont {I.}~\bibnamefont {{Gierz}}}, \bibinfo {author} {\bibfnamefont
  {M.~C.}\ \bibnamefont {{Hoffmann}}}, \bibinfo {author} {\bibfnamefont
  {M.}~\bibnamefont {{Le Tacon}}}, \bibinfo {author} {\bibfnamefont
  {T.}~\bibnamefont {{Loew}}}, \bibinfo {author} {\bibfnamefont
  {B.}~\bibnamefont {{Keimer}}}, \ and\ \bibinfo {author} {\bibfnamefont
  {A.}~\bibnamefont {{Cavalleri}}},\ }\bibfield  {title} {\enquote {\bibinfo
  {title} {Optically enhanced coherent transport in
  {YBa$_{2}$Cu$_{3}$O$_{6.5}$} by ultrafast redistribution of interlayer
  coupling},}\ }\href {\doibase 10.1038/nmat3963} {\bibfield  {journal}
  {\bibinfo  {journal} {Nat. Mater.}\ }\textbf {\bibinfo {volume} {13}},\
  \bibinfo {pages} {705--711} (\bibinfo {year} {2014})}\BibitemShut {NoStop}%
\bibitem [{\citenamefont {Mankowsky}\ \emph {et~al.}(2014)\citenamefont
  {Mankowsky}, \citenamefont {Subedi}, \citenamefont {Först}, \citenamefont
  {Mariager}, \citenamefont {Chollet}, \citenamefont {Lemke}, \citenamefont
  {Robinson}, \citenamefont {Glownia}, \citenamefont {Minitti}, \citenamefont
  {Frano}, \citenamefont {Fechner}, \citenamefont {Spaldin}, \citenamefont
  {Loew}, \citenamefont {Keimer}, \citenamefont {Georges},\ and\ \citenamefont
  {Cavalleri}}]{Mankowsky2014}%
  \BibitemOpen
  \bibfield  {author} {\bibinfo {author} {\bibfnamefont {R.}~\bibnamefont
  {Mankowsky}}, \bibinfo {author} {\bibfnamefont {A.}~\bibnamefont {Subedi}},
  \bibinfo {author} {\bibfnamefont {M.}~\bibnamefont {Först}}, \bibinfo
  {author} {\bibfnamefont {S.~O.}\ \bibnamefont {Mariager}}, \bibinfo {author}
  {\bibfnamefont {M.}~\bibnamefont {Chollet}}, \bibinfo {author} {\bibfnamefont
  {H.~T.}\ \bibnamefont {Lemke}}, \bibinfo {author} {\bibfnamefont {J.~S.}\
  \bibnamefont {Robinson}}, \bibinfo {author} {\bibfnamefont {J.~M.}\
  \bibnamefont {Glownia}}, \bibinfo {author} {\bibfnamefont {M.~P.}\
  \bibnamefont {Minitti}}, \bibinfo {author} {\bibfnamefont {A.}~\bibnamefont
  {Frano}}, \bibinfo {author} {\bibfnamefont {M.}~\bibnamefont {Fechner}},
  \bibinfo {author} {\bibfnamefont {N.~A.}\ \bibnamefont {Spaldin}}, \bibinfo
  {author} {\bibfnamefont {T.}~\bibnamefont {Loew}}, \bibinfo {author}
  {\bibfnamefont {B.}~\bibnamefont {Keimer}}, \bibinfo {author} {\bibfnamefont
  {A.}~\bibnamefont {Georges}}, \ and\ \bibinfo {author} {\bibfnamefont
  {A.}~\bibnamefont {Cavalleri}},\ }\bibfield  {title} {\enquote {\bibinfo
  {title} {Nonlinear lattice dynamics as a basis for enhanced superconductivity
  in $\mathrm{YBa}_2\mathrm{Cu}_3\mathrm{O}_{6.5}$},}\ }\href {\doibase
  10.1038/nature13875} {\bibfield  {journal} {\bibinfo  {journal} {Nature}\
  }\textbf {\bibinfo {volume} {516}},\ \bibinfo {pages} {71--73} (\bibinfo
  {year} {2014})}\BibitemShut {NoStop}%
\bibitem [{\citenamefont {Denny}\ \emph {et~al.}(2015)\citenamefont {Denny},
  \citenamefont {Clark}, \citenamefont {Laplace}, \citenamefont {Cavalleri},\
  and\ \citenamefont {Jaksch}}]{Denny2015}%
  \BibitemOpen
  \bibfield  {author} {\bibinfo {author} {\bibfnamefont {S.~J.}\ \bibnamefont
  {Denny}}, \bibinfo {author} {\bibfnamefont {S.~R.}\ \bibnamefont {Clark}},
  \bibinfo {author} {\bibfnamefont {Y.}~\bibnamefont {Laplace}}, \bibinfo
  {author} {\bibfnamefont {A.}~\bibnamefont {Cavalleri}}, \ and\ \bibinfo
  {author} {\bibfnamefont {D.}~\bibnamefont {Jaksch}},\ }\bibfield  {title}
  {\enquote {\bibinfo {title} {Proposed parametric cooling of bilayer cuprate
  superconductors by terahertz excitation},}\ }\href {\doibase
  10.1103/PhysRevLett.114.137001} {\bibfield  {journal} {\bibinfo  {journal}
  {Phys. Rev. Lett.}\ }\textbf {\bibinfo {volume} {114}},\ \bibinfo {pages}
  {137001} (\bibinfo {year} {2015})}\BibitemShut {NoStop}%
\bibitem [{\citenamefont {{Okamoto}}\ \emph {et~al.}(2016)\citenamefont
  {{Okamoto}}, \citenamefont {{Cavalleri}},\ and\ \citenamefont
  {{Mathey}}}]{Okamoto2016}%
  \BibitemOpen
  \bibfield  {author} {\bibinfo {author} {\bibfnamefont {J.-i.}\ \bibnamefont
  {{Okamoto}}}, \bibinfo {author} {\bibfnamefont {A.}~\bibnamefont
  {{Cavalleri}}}, \ and\ \bibinfo {author} {\bibfnamefont {L.}~\bibnamefont
  {{Mathey}}},\ }\bibfield  {title} {\enquote {\bibinfo {title} {Theory of
  enhanced interlayer tunneling in optically driven high-${T}_{c}$
  superconductors},}\ }\href {\doibase 10.1103/PhysRevLett.117.227001}
  {\bibfield  {journal} {\bibinfo  {journal} {Phys. Rev. Lett.}\ }\textbf
  {\bibinfo {volume} {117}},\ \bibinfo {pages} {227001} (\bibinfo {year}
  {2016})}\BibitemShut {NoStop}%
\bibitem [{\citenamefont {Homann}\ \emph {et~al.}(2021)\citenamefont {Homann},
  \citenamefont {Cosme}, \citenamefont {Okamoto},\ and\ \citenamefont
  {Mathey}}]{Homann2021}%
  \BibitemOpen
  \bibfield  {author} {\bibinfo {author} {\bibfnamefont {G.}~\bibnamefont
  {Homann}}, \bibinfo {author} {\bibfnamefont {J.~G.}\ \bibnamefont {Cosme}},
  \bibinfo {author} {\bibfnamefont {J.}~\bibnamefont {Okamoto}}, \ and\
  \bibinfo {author} {\bibfnamefont {L.}~\bibnamefont {Mathey}},\ }\bibfield
  {title} {\enquote {\bibinfo {title} {Higgs mode mediated enhancement of
  interlayer transport in high-${T}_{c}$ cuprate superconductors},}\ }\href
  {\doibase 10.1103/PhysRevB.103.224503} {\bibfield  {journal} {\bibinfo
  {journal} {Phys. Rev. B}\ }\textbf {\bibinfo {volume} {103}},\ \bibinfo
  {pages} {224503} (\bibinfo {year} {2021})}\BibitemShut {NoStop}%
\bibitem [{\citenamefont {Homann}\ \emph {et~al.}(2020)\citenamefont {Homann},
  \citenamefont {Cosme},\ and\ \citenamefont {Mathey}}]{Homann2020}%
  \BibitemOpen
  \bibfield  {author} {\bibinfo {author} {\bibfnamefont {G.}~\bibnamefont
  {Homann}}, \bibinfo {author} {\bibfnamefont {J.~G.}\ \bibnamefont {Cosme}}, \
  and\ \bibinfo {author} {\bibfnamefont {L.}~\bibnamefont {Mathey}},\
  }\bibfield  {title} {\enquote {\bibinfo {title} {Higgs time crystal in a
  high-${T}_{c}$ superconductor},}\ }\href {\doibase
  10.1103/PhysRevResearch.2.043214} {\bibfield  {journal} {\bibinfo  {journal}
  {Phys. Rev. Research}\ }\textbf {\bibinfo {volume} {2}},\ \bibinfo {pages}
  {043214} (\bibinfo {year} {2020})}\BibitemShut {NoStop}%
\bibitem [{\citenamefont {{Lin}}\ \emph {et~al.}(2021)\citenamefont {{Lin}},
  \citenamefont {{Rosa-Medina}}, \citenamefont {{Ferri}}, \citenamefont
  {{Finger}}, \citenamefont {{Kroeger}}, \citenamefont {{Donner}},
  \citenamefont {{Esslinger}},\ and\ \citenamefont
  {{Chitra}}}]{lin2021dissipationengineered}%
  \BibitemOpen
  \bibfield  {author} {\bibinfo {author} {\bibfnamefont {R.}~\bibnamefont
  {{Lin}}}, \bibinfo {author} {\bibfnamefont {R.}~\bibnamefont
  {{Rosa-Medina}}}, \bibinfo {author} {\bibfnamefont {F.}~\bibnamefont
  {{Ferri}}}, \bibinfo {author} {\bibfnamefont {F.}~\bibnamefont {{Finger}}},
  \bibinfo {author} {\bibfnamefont {K.}~\bibnamefont {{Kroeger}}}, \bibinfo
  {author} {\bibfnamefont {T.}~\bibnamefont {{Donner}}}, \bibinfo {author}
  {\bibfnamefont {T.}~\bibnamefont {{Esslinger}}}, \ and\ \bibinfo {author}
  {\bibfnamefont {R.}~\bibnamefont {{Chitra}}},\ }\href@noop {} {\enquote
  {\bibinfo {title} {Dissipation-engineered family of nearly dark states in
  many-body cavity-atom systems},}\ } (\bibinfo {year} {2021}),\ \Eprint
  {http://arxiv.org/abs/2109.00422} {arXiv:2109.00422 [cond-mat.quant-gas]}
  \BibitemShut {NoStop}%
\bibitem [{\citenamefont {Emary}\ and\ \citenamefont
  {Brandes}(2003)}]{Emary2003}%
  \BibitemOpen
  \bibfield  {author} {\bibinfo {author} {\bibfnamefont {C.}~\bibnamefont
  {Emary}}\ and\ \bibinfo {author} {\bibfnamefont {T.}~\bibnamefont
  {Brandes}},\ }\bibfield  {title} {\enquote {\bibinfo {title} {Chaos and the
  quantum phase transition in the dicke model},}\ }\href {\doibase
  10.1103/PhysRevE.67.066203} {\bibfield  {journal} {\bibinfo  {journal} {Phys.
  Rev. E}\ }\textbf {\bibinfo {volume} {67}},\ \bibinfo {pages} {066203}
  (\bibinfo {year} {2003})}\BibitemShut {NoStop}%
\bibitem [{\citenamefont {{Wagner}}(1975)}]{Wagner1975}%
  \BibitemOpen
  \bibfield  {author} {\bibinfo {author} {\bibfnamefont {M.}~\bibnamefont
  {{Wagner}}},\ }\bibfield  {title} {\enquote {\bibinfo {title} {{A nonlinear
  transformation of SU(3)-spin-operators to bosonic operators}},}\ }\href
  {\doibase 10.1016/0375-9601(75)90319-9} {\bibfield  {journal} {\bibinfo
  {journal} {Physics Letters A}\ }\textbf {\bibinfo {volume} {53}},\ \bibinfo
  {pages} {1--2} (\bibinfo {year} {1975})}\BibitemShut {NoStop}%
\end{thebibliography}%

\end{document}